\documentclass{jfm}
\usepackage{graphicx}
\usepackage{float}
\usepackage{rotating}
\usepackage[flushleft]{threeparttable}
\usepackage{epstopdf,epsfig}
\usepackage{lscape}
\usepackage{amsmath}
\usepackage{mathtools}
\usepackage{tabularx}
\usepackage{tikz}
\usepackage{centernot}
\usepackage{booktabs}
\usepackage{comment} 
\usepackage{subcaption}
\usepackage{siunitx}
\usepackage{lineno}
\usepackage{amsmath}
\usepackage{units}
\usepackage{nicefrac,xfrac}
\usepackage{soul}
\usepackage{ulem} 
\usetikzlibrary{shapes.geometric}
\usepackage{flafter}
\usepackage{hyphenat}
\usepackage{multicol}
\usepackage{multirow}
\usepackage{xr}
\usepackage{bm}

\usepackage{enumitem}
\newlist{myenumi}{description}{10}
\setlist[myenumi]{leftmargin=23pt,itemsep=2pt,topsep=2pt,parsep=2pt}
\newlist{myenumi2}{description}{10}
\setlist[myenumi2]{leftmargin=40pt,itemsep=2pt,topsep=2pt,parsep=2pt}

\usepackage{ragged2e}
\newcolumntype{P}[1]{>{\RaggedRight\hspace{0pt}}p{#1}}

\newcommand{\s}{\mathsf{s}}
\newcommand{\LL}{\textsf{L}}
\newcommand{\HH}{\textsf{H}}

\newcommand{\II}{\textsf{I}}
\newcommand{\TT}{\textsf{T}}

\renewcommand{\d}{\, \mbox{d}}

\newcommand{\uu}{\bm{u}}
\newcommand{\xx}{\bm{x}}

\newcommand{\Rigbb}{\overline{\overline{Ri_g}}}
\newcommand{\Sbb}{\bar{\bar{S}}}
\newcommand{\Nbb}{\bar{\bar{N}}}

\shorttitle{Experimental properties of stratified turbulence}
\shortauthor{A. Lefauve \& P. F. Linden}

\title{\Large{Experimental properties of continuously-forced, shear-driven,  stratified turbulence. \\ Part 2. Energetics, anisotropy, parameterisation.}}

\author{Adrien Lefauve and P. F. Linden}
\affiliation{Department of Applied Mathematics and Theoretical Physics, University of Cambridge \\ Centre for Mathematical Sciences, Wilberforce Road, Cambridge, CB3 0WA, UK.}

\begin{document}

\maketitle




\begin{abstract}

In this Part 2 we study further experimental properties of two-layer exchange flows in a stratified inclined duct (SID), which are turbulent, strongly-stratified, shear-driven, and continuously-forced. We analyse the same  state-of-the-art data sets using the same `core' shear layer methodology as in Part 1, but we focus here on turbulent energetics and mixing statistics. The detailed analysis of kinetic and scalar energy budgets reveals the specificity and scalings of SID turbulence, while energy spectra provide insight into the current strengths and limitations of our experimental data. The anisotropy of the flow at different scales characterises the turbulent kinetic energy production and dissipation mechanisms of Holmboe waves and turbulence. We then assess standard mixing parameterisations models relying on uniform eddy diffusivities, mixing lengths, flux parameters, buoyancy Reynolds numbers or turbulent Froude numbers, and we  compare representative values with the stratified mixing literature. The dependence of these measures of mixing on controllable flow parameters is also elucidated, providing asymptotic estimates that may be extrapolated to more strongly turbulent flows, quantified by the product of the tilt angle of the duct and the Reynolds number. These insights may serve as benchmark for the future generation of experimental data with superior spatio-temporal resolution required to probe increasingly vigorous turbulence.

\end{abstract}



\section{Introduction}\label{sec:intro}

In Part 1 we tackled a range of basic experimental properties of the continuously-forced, shear-driven, stratified turbulence generated by exchange flow in a stratified inclined duct (SID). We studied the permissible regions of the multi-dimensional parameter space, the mean flows and Reynolds-averaged dynamics, the gradient and equilibrium Richardson numbers, and the characterisation of turbulence with enstrophy and overturn volume fractions.

In this Part 2 we build on these results to tackle stratified turbulent energetics and mixing, perhaps the most enduring challenge in the community. In a recent review,  \cite{caulfield_open_2020} identified that there remain  ``leading-order open questions and areas of profound uncertainty'' to ``improv[e] community understanding, modeling, and parametrization of the subtle interplay among energy conversion pathways, turbulence, and irreversible mixing'' despite the ``proliferation of data obtained through direct observation, numerical simulation, and laboratory experimentation''. In another recent review,  \cite{gregg_mixing_2018} warned that ``We [...] do not know how relevant [idealized problems addressed by laboratory or numerical studies] are to ocean mixing'' and 
recommended that ``numerical and laboratory studies should help identify mixing mechanisms in the ocean with mimicking parameters that can be observed at sea, e.g., profiles of shear, stratification, turbulent dissipation and dissipation of scalar variance.'' 

Our motivations are that (i) the features of SID flows, highlighted in Part 1, allow them to mimick geophysically-relevant, shear-driven, stratified turbulence in some of its complexity; (ii) our 16 data sets of the density and three-component velocity fields in a three-dimensional volume, also introduced in Part 1, provide state-of-the-art access to the subtle energy pathways in `real' (experimentally-realisable) flows. In this paper we therefore undertake a comprehensive energetics analysis of these data sets, drawing on insights from previous studies of the SID (\citealp{meyer_stratified_2014}, hereafter ML14; \citealp{lefauve_regime_2019}, hereafter LPL19; and \citealp{lefauve_buoyancy_2020}, thereafter LL20) but using the same methodology and non-dimensional shear-layer framework as in Part 1, for more added value for the wider community.



The remainder of the paper is organised as follows. In \S~2 we introduce the background definitions and equations governing turbulent energetics in the SID. We will then make progress on the following sets of questions, to each of which we devote a section: 
\begin{myenumi}

\item[\textnormal{\S~3}\hspace{2.1ex}] How do the mean and turbulent kinetic energy and scalar variance vary across the Holmboe, intermittent and turbulent regimes? How do energy reservoirs and fluxes scale with respect to one another and with the flow parameters? What do their spectra reveal about these flows and about potential limitations of our measurements?
\item[\textnormal{\S~4}\hspace{2.1ex}] How anisotropic are  the velocity fields at larger and smaller scales? How does the shear-driven, stratified nature of Holmboe waves or turbulence affect the production and dissipation of turbulent kinetic energy?
\item[\textnormal{\S~5}\hspace{2.1ex}] How accurate are  `parameterisations' of stratified mixing using standard models such as eddy diffusivities or flux parameters? How do these quantities depend on key flow parameters? What does this tell us about the length scales of stratified turbulence in the SID? How to extrapolate our results to more strongly turbulent flows to inform future higher-resolution experiments?
\end{myenumi} 
Finally, we conclude in \S~6 and distill the key insights gained for the three-pronged (observational, numerical, experimental) modelling of stratified turbulence. 



\section{Background} \label{sec:energetics-equations}

In this section we give the background definitions and energy budget equations which form the basis of our energetics  analysis in \S\S~\ref{sec:energetics}-\ref{sec:param}.

\subsection{Definitions}

We first split the total local kinetic energy of the flow $K(\xx,t) \equiv (1/2) \uu \cdot \uu = \bar{K} + K'$ into a mean and a turbulent (or perturbation) kinetic energy, respectively, 
\begin{equation} \label{definition_Kbar_K'}
\bar{K}(y,z) \equiv \frac{1}{2} \bar{\uu} \cdot \bar{\uu} \ \ \ \textrm{and} \ \ \ K'(\xx,t) \equiv  \frac{1}{2} \uu' \cdot \uu' , 
\end{equation}
where we recall from Part 1 that the bar averages are $\bar{\cdot} \equiv \langle \cdot \rangle_{x,t}$, and the prime variables are perturbations with respect to these $x-t$ averages. 

By analogy, we also define the total scalar density variance $K_\rho \equiv (1/2) Ri_b^s \, \rho^2 = \bar{K}_\rho + K'_\rho$ into a mean and a turbulent (or perturbation) scalar variance, respectively,
\begin{equation} \label{definition_K'rho}
\bar{K}_\rho(y,z) \equiv \frac{1}{2} Ri_b^s \, \bar{\rho}^2 \ \ \ \textrm{and} \ \ \ K'_\rho(\xx,t) \equiv  \frac{1}{2} Ri_b^s \, \rho'^2.
\end{equation}
These variances are a useful and more convenient alternative to potential energies when estimating mixing. In particular $\bar{K}_\rho$ is more informative in SID flows than in most canonical stratified shear layers since the average density field $\bar{\rho}$ results entirely from mixing inside the duct, rather than being set as an initial condition. No mixing, i.e. the bimodal $\pm 1$ distribution from the external reservoirs, corresponds to a maximum $(1/Ri_b^s)\langle \bar{K}_\rho \rangle = 1/2$ (and $K'_\rho=0$).  By contrast, complete mixing (uniform $\bar{\rho}=0$)  corresponds to a minimum $(1/Ri_b^s) \langle \bar{K}_\rho \rangle = 0$, and a linear stratification with uniform gradient across the shear layer $\p_z \bar{\rho} =-1$ corresponds to an intermediate value of $(1/Ri_b^s)\langle \bar{K}_\rho \rangle = 1/3$.

\subsection{Evolution equations}

The averaged equations of $\bar{K}, \bar{K}_\rho,$ and the temporal evolution equations of $K',K'_\rho$ follow from the equations of motion (3.5) in Part 1: 
\begin{subeqnarray} \label{dKdt}
\overline{\p_t K}(y,z) &=&  \Phi^{\bar{K}} \, -  \mathcal{P} + \mathcal{F} - \bar{\epsilon}, \slabel{dKbardt} \\
\p_t K'(\xx,t) &=&  \Phi^{K'}  + \mathcal{P} - \mathcal{B} -  \mathcal{E}, \slabel{dK'dt} \\
\overline{ \p_t K_\rho}(y,z) &=&  \Phi^{\bar{K}_\rho}   -  \mathcal{P}_\rho,    \slabel{dKbarrhodt}\\
\p_t K'_\rho (\xx,t) &=&  \Phi^{K'_\rho }   +  \mathcal{P}_\rho  \  \ \ \ -   \chi  \slabel{dK'rhodt},
\end{subeqnarray}
where the mean temporal gradients have the form $\overline{\p_t K}= (1/L_t)(\langle K \rangle_x(t=L_t)-\langle K \rangle_x(t=0)) \approx 0$ in quasi-steady state (similarly for $\bar{K}_\rho$). All $\Phi$ terms are transport terms that will be discussed in \S~\ref{sec:energetics-boundary-fluxes}.

The mean kinetic energy equation \eqref{dKbardt} has three source/sink terms: the production of turbulent kinetic energy $\mathcal{P}$ (generally positive) by interaction of the off-diagonal (deviatoric) Reynolds stresses with the mean shear, the gravitational forcing term $\mathcal{F}$ (generally positive) transferring energy from the mean potential energy (not shown here), and the viscous dissipation of the mean $\bar{\epsilon}$ (always positive):
\begin{equation} \label{dKdt-terms-1}
 \mathcal{P} \equiv - \overline{u'v'} \p_y \bar{u} -  \overline{u'w'}  \p_z \bar{u}, \qquad \mathcal{F}  \equiv  Ri_b^s \, \sin \theta \, \bar{u}\bar{\rho} ,  \qquad 
\bar{\epsilon} \equiv  \frac{2}{Re^s} \bar{s}_{ij} \bar{s}_{ij}>0,
\end{equation}
where the mean strain rate tensor is $\bar{s}_{ij} \equiv (\p_{x_i} \bar{u}_j  + \p_{x_j} \bar{u}_i)/2$ and we  implicitly sum over repeated indices (unless specified otherwise).

The remaining equations \eqref{dK'dt}-\eqref{dK'rhodt} have four further volumetric terms: the turbulent buoyancy flux $\mathcal{B}$ (transferring energy to the turbulent kinetic energy, generally positive), the production of turbulent scalar variance $\mathcal{P}_\rho$ (generally positive), the turbulent dissipation $\mathcal{E}$ (always positive), and the turbulent scalar dissipation $\chi$ (always positive):
\begin{equation} \label{dKdt-terms-2}
\mathcal{B}  \equiv  Ri_b^s \ \overline{w'\rho'}, \quad \mathcal{P}_\rho \equiv - Ri_b^s \, \overline{w' \rho'} \p_z \bar{\rho}, \quad  
\mathcal{E} \equiv  \frac{2}{Re^s} s'_{ij} s'_{ij}>0, \quad \chi \equiv \frac{Ri_b^s}{Re^s \, Pr} \p_{x_j} \rho'\p_{x_j} \rho'>0,
\end{equation}
where  $s'_{ij} \equiv (\p_{x_i} u'_j  + \p_{x_j} u'_i)/2$. All terms in \eqref{dKdt-terms-1}-\eqref{dKdt-terms-2} are functions of $y,z$ only, except for $\mathcal{E}$ and $\chi$, which are function of $\xx,t$.

We see in \eqref{dKdt-terms-2} that $\mathcal{P}_\rho$ is proportional to $\mathcal{B}$ in the simple case of linear stratification. Moreover, $\mathcal{P}_\rho=\mathcal{B}$ if $\p_z \bar{\rho}=-1$ (linear mixing layer spanning the entire shear layer),  since in this case $\mathcal{B}$ is a source term for the turbulent potential energy, which is exactly equal to $K'_\rho$ (as noted by \citealp{taylor_testing_2019}, \S~3).

\subsection{Approximations}

A few simplifying approximations were made in \eqref{dKdt}-\eqref{dKdt-terms-2}. First, in \eqref{dK'rhodt} we neglected the molecular scalar dissipation $Ri_b^s/(Re^s  Pr) (\p_{x_j} \bar{\rho}\p_{x_j} \bar{\rho}) \approx 0$ (requiring $|\p_{x_j} \bar{\rho}|\ll \sqrt{Re^s Pr/Ri_b^s}$ which is true here for $Pr=700$). Second, in the definition of $\mathcal{P}$ we assumed parallel mean flow, i.e. $\bar{v},\bar{w} \approx 0$, and in $\mathcal{P}_\rho$ we assumed $\p_y\bar{\rho} \approx 0$ (which are good approximations). Third, in $\mathcal{F}$ we assumed no mean vertical buoyancy flux, i.e. $\bar{w} \bar{\rho} \approx 0$ (this term is key in horizontal exchange flows at $\theta=0$, but negligible in long ducts at $\theta>0$  since the mean slope of the density interface is small, as explained in LPL19, \S~4.3). Fourth, in $\mathcal{B}$ we assumed $\overline{u'\rho'},\overline{v'\rho'} \approx 0$ and $\cos \theta \approx 1$.

\subsection{Boundary fluxes} \label{sec:energetics-boundary-fluxes}

The transport terms $\Phi$ in \eqref{dKdt} represent the divergence of advective, pressure and viscous/molecular fluxes:
\begin{equation}\label{K-transport-terms}
\begin{gathered}  
\Phi^{\bar{K}} \equiv  - \overline{\p_x (u K)} - \overline{\p_x (u p)} + \frac{2}{Re^s} \overline{\p_x(u s_{11}}), \quad
\Phi^{K'}  \equiv   \p_{x_i} \big( - u'_i K' - u'_i p' + \frac{2}{Re^s} u'_j  s'_{ij} \big), \\
\Phi^{\bar{K}_\rho} \equiv -  \overline{\p_x(uK_\rho)}, \qquad \Phi^{K'_\rho} \equiv \p_{x_i} \big(- u'_i K'_\rho + \frac{1}{Re^s \, Pr} \p_{x_i} K'_\rho\big),
\end{gathered}
\end{equation}
%
where, as above, we assumed parallel mean flow and negligible molecular transport in $\Phi^{\bar{K}}$ and $\Phi^{\bar{K}_\rho}$, and where $\overline{\p_x \phi}$ denotes the mean gradient along $x$ (non-zero if $\phi$ is non-periodic). When averaged over a volume, these divergence terms become boundary fluxes. 

These boundary fluxes are typically neglected in the stratified turbulence literature, because they usually conveniently vanish in idealised geometries (e.g. for periodic boundary conditions), greatly simplifying \eqref{dKdt}. In the SID geometry, they are unfortunately slightly more complicated as we explain below.

In the $y$ and $z$ directions, $\Phi^{K'}$ and $\Phi^{K'_\rho}$ will not generally cancel if the volume-average is done over the shear layer  (as in this paper) because the boundaries do not include duct walls (whereas LPL19, \S~4.2.1 included them). In other words, turbulent fluctuations can in principle be transported freely across our  shear layer `imaginary' boundary ($y=\pm L_y, z=\pm 1$) to (or, more rarely, from) the near-wall region. 

More importantly, in the $x$ direction, most boundary fluxes can generally be neglected when $\theta \gtrsim \arctan A^{-1}$, where $A\equiv L/H$ is the length-to-height aspect ratio of the duct (high in the long ducts of interest here, $A=30$ in our set-up). In these so-called \emph{forced flows}, the mean slope of the density interface is small and the flow is approximately periodic (see LPL19, \S~4.3 and their Appendix~B). This applies in particular to $-\overline{\p_x (uK)}$, which is important and $<0$ when $\theta\approx 0$, but unimportant and $\approx 0$ in forced flows.  

We, however, note two exceptions. First, our Part 1 results on the unexpected nature of the estimated mean pressure gradient $\Pi = -\overline{\p_x p}$  (weakening $\bar{u}$ rather than strengthening it) suggest that the simple hydrostatic pressure assumed in LPL19's Appendix~B may not be correct and, consequently, that $-\overline{\p_x (u p)}$ may not be neglected. However, for simplicity, and due to our inability to measure it directly, we ignore it in this paper until future work sheds light on it. Second, the flux of mean scalar variance $\Phi^{\bar{K}_\rho}>0$ represents the continuous inflow of unmixed fluid from the reservoirs, countering the effects of mixing, and must be retained to ensure that a steady state for $\bar{K}_\rho$ is possible.

\subsection{Steady-state balances} \label{sec:energetics-steady}

In our unsteady flows, steady state  ($\p_t = 0$) cannot be expected in the pointwise and instantaneous sense of \eqref{dKdt}. It can however be expected in a time- and volume-averaged sense, leading to the following balances:
\begin{subeqnarray} \label{dKdt_0_prelim}
0 &\approx& \langle \mathcal{F}\rangle - \langle \mathcal{P} \rangle - \langle \bar{\epsilon} \rangle \slabel{dKdt_0_prelim-1} \\ 
0 &\approx& \langle \mathcal{P}\rangle - \langle \mathcal{B} \rangle - \langle \mathcal{E} \rangle \slabel{dKdt_0_prelim-2}\\ 
0 &\approx& \langle \Phi^{\bar{K}_\rho} \rangle - \langle \mathcal{P}_\rho \rangle ,\slabel{dKdt_0_prelim-3}\\
0 &\approx&  \langle \mathcal{P}_\rho \rangle -   \langle \chi \rangle,\slabel{dKdt_0_prelim-4}
\end{subeqnarray}
where we recall from  Part 1  that $\langle \cdot \rangle \equiv \langle \cdot \rangle_{x,y,z,t}$. In the above, we assumed  for simplicity that all boundary fluxes were negligible, except the essential $\Phi^{\bar{K}_\rho}$  sustaining the steady-state scalar dissipation. We also assumed that all mean temporal gradients are negligible $\langle \p_t \phi \rangle_t = (1/L_t)(\phi(L_t)-\phi(0)) \approx 0$ (verified in our data).

These above balances can alternatively be expressed as two `independent' estimations of the mean turbulent dissipation rates $\langle\mathcal{E}\rangle,\langle \chi\rangle$:
\begin{subeqnarray} \label{dKdt_0}
\langle \mathcal{E} \rangle &\approx& \langle \mathcal{P}\rangle - \langle \mathcal{B} \rangle  \slabel{dKdt_0-1}  \\ 
&\approx& \langle \mathcal{F}\rangle - \langle\mathcal{B}\rangle - \langle \bar{\epsilon}\rangle , \slabel{dKdt_0-2} \\
   \langle \chi \rangle &\approx& \langle \mathcal{P}_\rho \rangle  \slabel{dKdt_0-3} \\ 
   &\approx& \langle \Phi^{\bar{K}_\rho} \rangle \slabel{dKdt_0-4}.
\end{subeqnarray}
Equation \eqref{dKdt_0-1} represents the classical balance of \cite{osborn_estimates_1980}, \eqref{dKdt_0-3} represents the classical balance of \cite{osborn_oceanic_1972}, while  \eqref{dKdt_0-2} and \eqref{dKdt_0-4} are more specific to SID flows.

\section{Energetics} \label{sec:energetics}

We now use experimental data to test the validity of the above equations and approximations, and obtain further insight into the time- and volume-averaged energy reservoirs and their fluxes in \S~\ref{sec:energetics-time-vol-avgs}, their spatio-temporal structures in \S~\ref{sec:energetics-spatiotemp}, their spectra in \S~\ref{sec:energetics-spectra}, and the limitations in their accuracy in \S~\ref{sec:energetics-spectra-errors}.

\subsection{Time and volume averages}\label{sec:energetics-time-vol-avgs}

\subsubsection{Energy reservoirs} \label{sec:energetics-mean-reservoirs}

\begin{figure}
\centering
\includegraphics[width=1.05\textwidth]{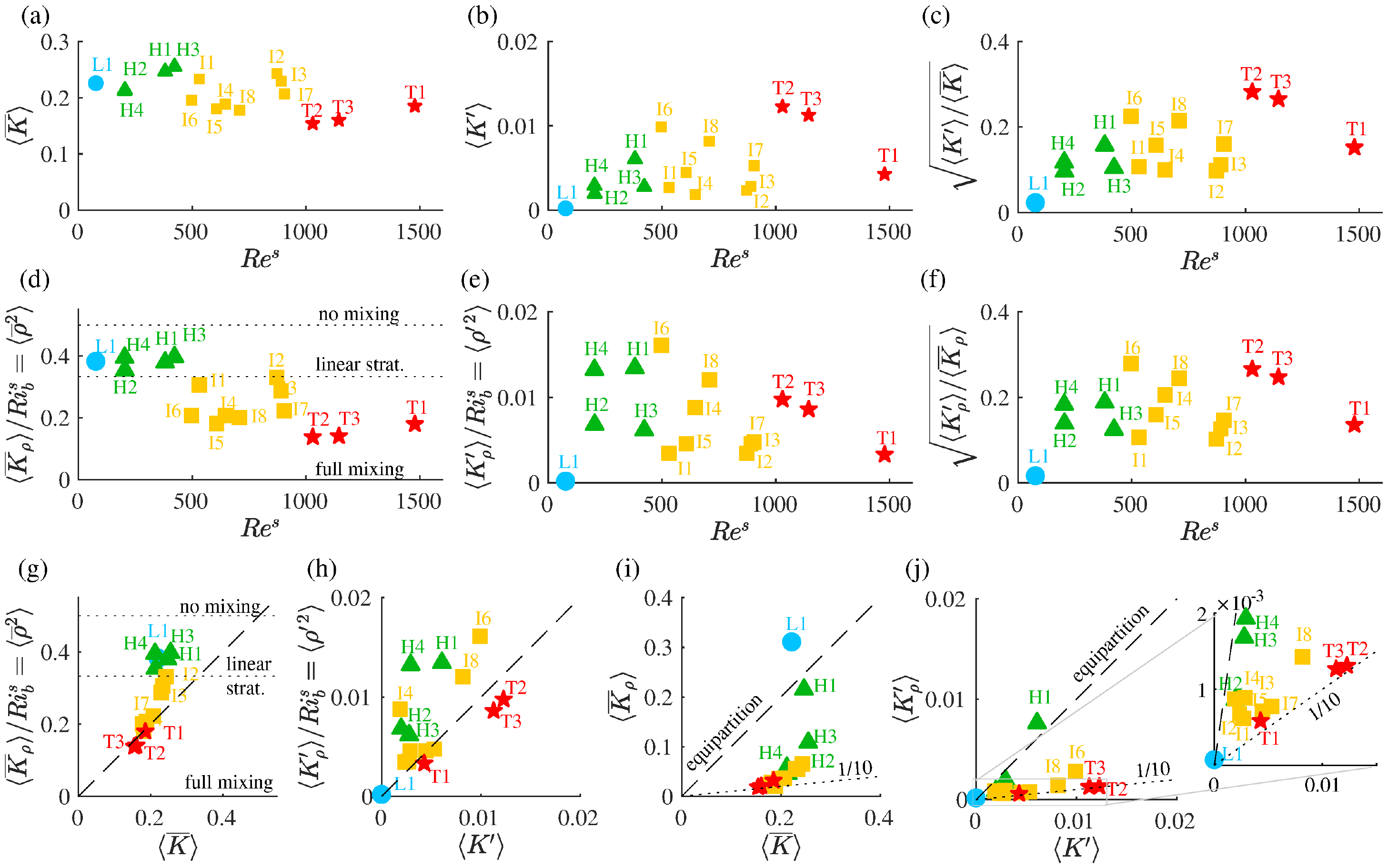}
    \caption{Steady-state energy reservoirs in all 16 data sets. \emph{(a-c)} Mean  and turbulent  kinetic energies, and their square-root ratio, as function of $Re^s$ (separating flow regimes).  \emph{(d-f)} Mean  and turbulent  scalar variances and their square-root ratio (rescaled from \eqref{definition_Kbar_K'} to obtain $\langle \bar{\rho}^2 \rangle$, $\langle \rho'^2 \rangle$).  \emph{(g-j)} Correlation between scalar variance and kinetic energies (in \emph{i-j} we show the full $\bar{K}_\rho, K'_\rho$ to test for potential-to-kinetic energy partitions). Symbol shapes and colours follow the flow regimes, as in Part 1. All dashed lines have slope one. Dotted lines are labelled explicitly.}
    \label{fig:energetics_mean_values}
\end{figure}

In figure~\ref{fig:energetics_mean_values} we plot the steady-state energy reservoirs in all 16 data sets, both as function of $Re^s$ (panels~\emph{a-f}, top two rows), and as correlation plots (panels~\emph{g-j}, bottom row).

Note that our definition of turbulent perturbations around the $x-t$ mean flow can attribute artificially high energies to  $\LL$ and $\HH$ flows, whose perturbations $u'\equiv u - \bar{u}$ and $\rho' = \rho-\bar{\rho}$ can exhibit slight residual $x-t$ structure due to the nature of our exchange flow (slightly non-parallel in $x$ and/or accelerating or decelerating in $t$). Therefore, in figure~\ref{fig:energetics_mean_values} we removed this artefact (not due to the turbulent or wave motions or interest) by subtracting from $\langle K'\rangle$, $\langle K'_\rho\rangle$ the mean $x-t$ variance corresponding to the zero $x$-wavenumber and temporal frequency content of their respective spectra (i.e. we subtracted from $\langle u'^2\rangle_{x,y,z,t}$, the components $\langle \langle u'\rangle^2_x\rangle_{y,z,t}$ and $\langle \langle u'\rangle^2_t\rangle_{x,y,z}$ and similarly for $\rho'$). We verified that $\II$ and $\TT$ flows are almost unaffected by this correction. We return to this  in our discussion of energy spectra in \S~\ref{sec:energetics-spectra} and  Appendix~\ref{sec:Appendix-spectra-energy-k0}.

The mean kinetic energy $\langle \bar{K} \rangle$ (panel~\emph{a}) is approximately constant around $0.2$ in all flows, with values decreasing from $0.25$ in $\LL, \HH$ flows to $0.15$ in $\TT$ flows. The turbulent kinetic energy $\langle K' \rangle$ (panel~\emph{b}) increases from $0$ in $\LL$ flows to around $0.01$ in $\TT$ flows (e.g. T2, T3). The square-root ratio of turbulent-to-mean kinetic energies $\sqrt{\langle K' \rangle /\langle \bar{K} \rangle}$ (panel~\emph{c}), indicating the relative magnitude of velocity fluctuations, is $10-15$~\% in $\HH$ flows, and $10-20$~\% in $\II$ flows (with significant spread) and up to $25$~\% in $\TT$ flows. 

We now turn to the scalar variance reservoirs. Although $\bar{K}_\rho, K'_\rho$ are preferred when discussing energy fluxes (as in \S~\ref{sec:energetics-equations}) because of their interpretation as proxy for potential energy under linear stratification, we first consider in panels~\emph{d-h} the rescaled quantities $\langle \bar{K}_\rho \rangle/ Ri_b^s \equiv \langle \bar{\rho}^2 \rangle$ and $\langle K'_\rho \rangle/Ri_b^s \equiv \langle \rho'^2 \rangle$, which are more straightforward measures of scalar variance. High values of mean variance $\langle \bar{\rho}^2 \rangle \approx 0.4$ (panel~\emph{d}) confirm that very little mixing takes place in $\LL$ and $\HH$ flows beyond molecular diffusion (close to the no-mixing upper bound of $0.5$, see dotted lines). Mixing increases in $\II$ flows, where an intermediate layer of approximately uniform density achieves `more' mixing than a uniformly linear stratification ($\langle \bar{\rho}^2 \rangle <1/3$, see dotted lines), while $\TT$ flows are halfway between linear and full mixing ($\langle \bar{\rho}^2 \rangle \approx 1/6$). The turbulent variance $\langle \rho'^2 \rangle$ (panel~\emph{e}) is, surprisingly, higher in some $\HH$ flows than in most $\II$ and $\TT$ flows. This reflects the fact that Holmboe waves on a sharp interface can generate very large perturbations on either side of it (due to high $|\rho'|$ values), compared to a well-mixed turbulent layer (low $|\rho|$ values). This effect partially disappears when considering the relative square-root of turbulent-to-mean variance $\sqrt{\langle K'_\rho \rangle /\langle \bar{K}_\rho \rangle} = \sqrt{\langle \rho'^2 \rangle /\langle \bar{\rho}^2 \rangle}$ (panel~\emph{f}), typically higher in $\II$ and $\TT$ flows, and reaching a maximum of $25$~\%, just as in kinetic energies (panel~\emph{c}).

We further see that the mean scalar variance $\langle \bar{\rho}^2 \rangle$ is closely correlated to the mean kinetic energy $\langle \bar{K} \rangle$ (panel~\emph{g}), especially in $\II$ and $\TT$ flows, where they become equal (dashed line). This reflects our observation in Part 1 of self-similarity $\langle \bar{u} \rangle_y(z) \approx \langle \bar{\rho} \rangle_y(z)$ in $\TT$ flows, i.e. that momentum and density become equally mixed. This general correlation  in $\II$ and $\TT$ flows also extends to the turbulent energies $\langle \rho'^2 \rangle$ and $\langle K' \rangle$ (panel~\emph{h}).

We now turn to the potential-to-kinetic energy partitions. The mean partition $\langle \bar{K}_\rho \rangle/\langle \bar{K} \rangle$ (panel~\emph{i}) drops from  $\approx 1$ (equipartition, see dashed line) in L1 and H1 (where $Ri_b^s \approx 0.5-1$) down to $\approx 0.1$ (1/10 partition, see dotted line) in the late $\II$ and $\TT$ regimes (where $Ri_b^s \approx 0.1-0.2$). The turbulent partition $\langle K'_\rho \rangle / \langle K' \rangle$ (panel~\emph{j}) follows a similar trend of equipartition in all $\HH$ flows, and asymptotic $1/10$ partition in $\TT$ flows (see the zoomed-in inset for more details).

\subsubsection{Energy fluxes} \label{sec:energetics-mean-fluxes}

\begin{figure}
\centering
\includegraphics[width=0.9\textwidth]{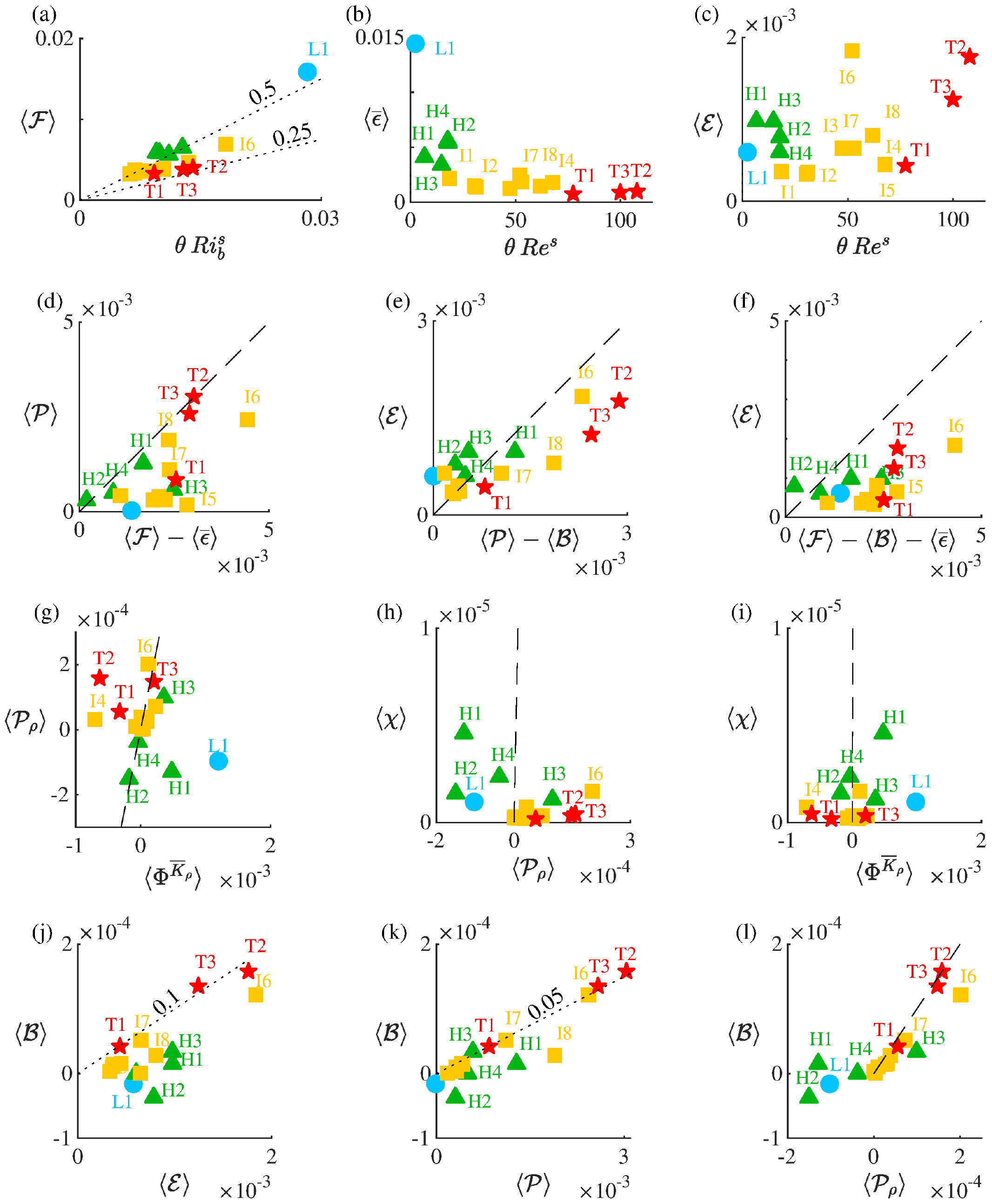}
    \caption{Steady-state energy fluxes in all 16 data sets. \emph{(a)} Mean kinetic energy forcing (power source) as function of $\theta Ri^s_b$; \emph{(b-c)} mean kinetic energy dissipation, and turbulent kinetic energy dissipation (power sinks) as function of  $\theta Re^s$ ($\approx \theta Re^h$ identified as proxy for regime transitions in LPL19). \emph{(d-f)} Test of the approximate kinetic balances \eqref{dKdt_0_prelim-1},  \eqref{dKdt_0-1}, \eqref{dKdt_0-2}, respectively. \emph{(g-i)} Test of the approximate scalar variance balances \eqref{dKdt_0_prelim-3},  \eqref{dKdt_0-3}, \eqref{dKdt_0-4}, respectively. \emph{(j-l)} Test of three further commonly-used ratios: $\langle \mathcal{B} \rangle  / \langle \mathcal{E} \rangle  \equiv \Gamma$, $\langle \mathcal{B} \rangle  / \langle \mathcal{P} \rangle  \equiv Ri_f$, and $\langle \mathcal{B} \rangle  / \langle \mathcal{P}_\rho \rangle$ ($\equiv 1$ when $\p_z \bar{\rho}=1$), respectively. All dashed lines have slope 1 and denote expected equality between fluxes. Dotted lines are labelled explicitly.}
    \label{fig:energetics_mean_fluxes}
\end{figure}

In figure~\ref{fig:energetics_mean_fluxes} we plot the steady-state energy fluxes in all 16 data sets. All gradients were computed using second-order accurate finite differences.

In panels \emph{a-c} (top row) we investigate the dependence of the kinetic energy source $\langle \mathcal{F} \rangle$ and sinks $ \langle \bar{\epsilon} \rangle, \langle \mathcal{E}\rangle$ with respect to two key groups of parameters $\theta Ri_b^s $ and $\theta Re^s$, respectively. Note that $\theta$ is in radians, and recall from Part 1 (see figure 2) how these output parameters depended on input parameters:  $Ri_b^s \propto \theta^{-0.9}(Re^h)^{-0.4}$ and $Re^s \propto \theta^{0.7}(Re^h)^{1.4}$. As expected from its definition \eqref{dKdt-terms-1}, $\langle \mathcal{F}\rangle \propto \theta Ri_b^s$, with a factor $\approx 0.5$ in $\LL$ and $\HH$ flows, decreasing to $\approx 0.25$ in $\TT$ flows (due to a lower $\langle \bar{u}\bar{\rho}\rangle$). The mean dissipation $\langle \bar{\epsilon}\rangle$ dominates over the turbulent dissipation $\langle \mathcal{E} \rangle$ at low $\theta Re^s$ ($\LL$ and $\HH$ flows), but decreases to become comparable or lower at higher $\theta Re^s= O(100)$ (T2 and T3). These observations in panels \emph{a-c} are key -- and almost defining -- features of SID flows: hydraulic control of two-layer exchange flows sets an upper bound on the magnitude of the mean flow (set by the dimensional scale $\sqrt{g'H}$ and thus $\langle |\bar{u}| \rangle \lesssim 1/2$ or $\langle \bar{K} \rangle \lesssim 1/4$), causing a plateau in $\langle \bar{\epsilon}\rangle$ in the $\II$/$\TT$ regimes, and thus an increase in $\langle \mathcal{E} \rangle$, which eventually dominates to match the increased $\langle \mathcal{F} \rangle$ at higher $\theta$ (see ML14 and LPL19).

In panels \emph{d-f} (second row) we test the approximate kinetic energy balances of \eqref{dKdt_0_prelim-1},  \eqref{dKdt_0-1}, \eqref{dKdt_0-2}, respectively. The mean balance $\langle \mathcal{P} \rangle  \approx \langle \mathcal{F} \rangle  - \langle \bar{\epsilon} \rangle$ is only verified (dashed line) in a subset of flows (e.g. H1, H2, H4 I8, T2, T3). The systematic underestimation of $\langle \mathcal{P} \rangle$ is due partly to the neglected boundary flux $\langle \Phi^{\bar{K}} \rangle$, and partly to our limited resolution of small-scales fluctuations (which are  needed to measure  $\langle \mathcal{P} \rangle$  but not $\langle \mathcal{F} \rangle$ and $\langle \bar{\epsilon} \rangle$). The turbulent balance of \cite{osborn_estimates_1980} $\langle \mathcal{E} \rangle  \approx \langle \mathcal{P} \rangle  - \langle \mathcal{B} \rangle$ is also verified in a (different) subset of flows. The general underestimation of $\langle \mathcal{E} \rangle$, especially in $\II$ and $\TT$ flows, is primarily due to the limited resolution of gradients of small-scale velocity fluctuations (needed to measure $\langle \mathcal{E} \rangle$ but not $\langle \mathcal{P} \rangle$ and  $\langle \mathcal{B} \rangle$). The balance $\langle \mathcal{E} \rangle  \approx \langle \mathcal{F} \rangle  - \langle \mathcal{B} \rangle -  \langle \bar{\epsilon} \rangle$ follows from the previous two balances, and is thus the most poorly-verified overall.

In panels \emph{g-i} (third row), we test the approximate scalar variance balances \eqref{dKdt_0_prelim-3},  \eqref{dKdt_0-3}, \eqref{dKdt_0-4}, respectively. The balance between production of turbulent variance and advective flux of mean variance (from unmixed fluid coming into the domain) $\langle \mathcal{P}_\rho \rangle  \approx \langle \Phi^{\bar{K}_\rho} \rangle$ (panel~\emph{g}) is verified in most flows (e.g. H2, H4, T3 and most $\II$ flows except I4), although the cluster near $0$ is inconclusive.  Some $\HH$ flows (H2 and H4) even show equality between negative values, which suggests that: (i)  the net effect of Holmboe wave turbulence in the measurement volume is to increase (rather than decrease) scalar variance, by sharpening (rather than broadening) the mean density interface, consistent with the findings of \cite{zhou_diapycnal_2017,salehipour_turbulent_2016} and our Reynolds-averaged profiles in Part 1; and/or (ii) this sharpening must be countering the net advection of mixed fluid into the volume, which means that mixing must take place outside the length of the duct occupied by Holmboe waves, presumably near the ends of the duct where plumes discharge turbulently into the reservoirs and interact with the incoming fluid, entraining mixed fluid back into the duct. Negative values of $\langle \Phi^{\bar{K}_\rho} \rangle<0$ in I4, T1 and T2 are, however, surprising and likely the result of experimental noise in the computation of this mean gradient.
The turbulent balance of \cite{osborn_oceanic_1972} $\langle \mathcal{\chi} \rangle  \approx \langle\mathcal{P}_\rho \rangle>0$ (panels \emph{h}), only valid for broadening-type ($\II$ and $\TT$) flows (because of the neglect of $\langle \Phi^{K'_\rho} \rangle$), is far from being verified even in these flows. The systematic and severe underestimation of $\langle \mathcal{\chi} \rangle$ is due to our severely limited resolution of small-scale density gradients (more severe than for $\langle \mathcal{E} \rangle$, because  $\rho'$ contains energetic length scales that are approximately a factor $\sqrt{Pr} \approx 25$ smaller than $\uu'$). Finally, the balance $\langle \mathcal{\chi} \rangle  \approx  \langle \Phi^{\bar{K}_\rho} \rangle>0$ (panels \emph{i}) follows from the previous two balances and is thus equally poorly verified. We explain the reasons for these limitations in  \S~\ref{sec:energetics-spectra-errors}.

In panels \emph{j-l}, we test the correlation of $\langle \mathcal{B} \rangle$ with the three other turbulent fluxes $\langle \mathcal{E} \rangle, \langle \mathcal{P} \rangle, \langle \mathcal{P}_\rho \rangle$ respectively, in order to assess the relevance and numerical value of the following ratios:
\begin{equation} \label{def-Gamma-Rf}
   \frac{ \langle \mathcal{B} \rangle}{ \langle \mathcal{E} \rangle } \equiv \Gamma, \quad \frac{\langle \mathcal{B} \rangle}{   \langle \mathcal{P} \rangle}  \equiv R_f, \quad \frac{\langle \mathcal{B} \rangle}{   \langle \mathcal{P}_\rho \rangle} \equiv 1 \  \text{when} \  \p_z \bar{\rho}=-1 \  \text{everywhere}.
\end{equation}
The flux parameter $\Gamma$ and the flux Richardson number $R_f$ date back to \cite{osborn_estimates_1980} and have been extensively used in the literature to parameterise the `taxation rate' of stratification on turbulent dissipation \citep{caulfield_open_2020}. Although often assumed constant, dimensional analysis suggests that $\Gamma$ and $R_f$ are functions $(\theta,Re^s,Ri^s_b,R,Pr)$ until proven otherwise. First, our data show that $\langle \mathcal{B} \rangle \propto \langle \mathcal{E} \rangle$ only in late $\II$ flows and in all $\TT$ flows (panel~\emph{j}), where the slope indicates an asymptotic ratio $\Gamma \approx 0.1$ (dotted line), about half the commonly-used value of 0.2 in the literature. The slightly negative values of $\langle \mathcal{B} \rangle$ can be explained by the slight non-periodicity of exchange flows at low tilt angles $0<\theta \lesssim \arctan A^{-1} \approx 1/30 \approx 2^\circ$: the convective acceleration of each layer ($u'\p_x u'>0$) caused by a tilting interface produces downward flow $(w'<0)$ in the dense layer ($\rho'>0$) and \emph{vice versa}, resulting in a net volume-averaged $\langle \mathcal{B}\rangle = Ri_b^s \langle w'\rho'\rangle <0$ in the absence of turbulence. This effect vanishes in more turbulent flows at larger tilt angles, where we instead tend to slightly overestimate $\Gamma$ by our underestimation of its denominator $\langle \mathcal{E} \rangle$ (compared to its numerator $\langle \mathcal{B} \rangle$, due to limitations in our computation of small-scale gradients).  Second, we see that $\langle \mathcal{B} \rangle \propto \langle \mathcal{P} \rangle$ in most $\II$ and $\TT$ flows (panel~\emph{k}), where the slope indicates an asymptotic ratio $R_f \approx 0.05$ (dotted line), about a third of the commonly-used value of 0.15 in the literature. Third, we see that $\langle \mathcal{B} \rangle \approx   \langle \mathcal{P}_\rho \rangle$ (dashed line) in most $\II$ and $\TT$ flows (panel~\emph{l}), which is consistent with the theory under linear stratification (where $\p_z \bar{\rho}=-1$), despite such a stratification being only achieved approximately in T3 (see Part 1, figure~3\emph{p}). We return to these parameters in more detail in \S~\ref{sec:param}.





\subsubsection{Estimations of $\langle \mathcal{E}\rangle$ and $\langle \chi\rangle$ from non-dimensional parameters} \label{sec:energetics-estimations-E-chi}

In this section we combine the steady-state energy balances of \S~\ref{sec:energetics-steady} and the experimental results of \S~\ref{sec:energetics-mean-fluxes} to propose indirect estimations (or proxies) of $\langle \mathcal{E}\rangle$ and $\langle \chi\rangle$ that are  insightful and more accurate than their direct computations, which rely on small-scale gradients.

From \eqref{dKdt_0-1} and \eqref{def-Gamma-Rf}, we take advantage of the fact that $\mathcal{P}$ is measured with better accuracy than $\mathcal{E}$ to propose
\begin{subeqnarray}
   \langle \mathcal{E}\rangle &\approx& \langle \mathcal{P}\rangle - \langle \mathcal{B}\rangle \\
   &\approx&  (1-R_f)  \langle\mathcal{P}\rangle \\
   &\approx&  \frac{1}{1+\Gamma} \langle \mathcal{P}\rangle,
\end{subeqnarray}
which means that
\begin{equation}\label{eq-link-gamma-Rf}
R_f \approx \frac{\Gamma}{1 + \Gamma} \ \  \ \text{or} \  \ \ \Gamma \approx \frac{R_f}{1 - R_f}.   
\end{equation}
These estimations depend on the balance \eqref{dKdt_0-1} and the assumption \eqref{def-Gamma-Rf} that the fluxes $\langle \mathcal{E}\rangle, \langle \mathcal{B}\rangle , \langle \mathcal{P}\rangle$ are proportional to one another,  approximately verified in $\TT$ flows. Note however that our measurements gave slightly incompatible values of $\Gamma \approx 0.1$ and $R_f \approx 0.05$. We believe that $\Gamma$ is overestimated due to the underestimation of its denominator $\langle \mathcal{E}\rangle$, and that values of $\Gamma \approx 0.05/0.95 \approx 0.05$ are more realistic.

From \eqref{dKdt_0-2}, we take advantage of the fact that $\mathcal{F}$ is measured with even better accuracy than $\mathcal{P}$ to propose a series of further approximations of $\langle \mathcal{E}\rangle$ valid in the limit of very turbulent flows ($\theta Re^s \gg 100$): 
\begin{subeqnarray}\label{estim-E-all}
   \langle \mathcal{E}\rangle &\approx& \langle \mathcal{F}\rangle - \langle \mathcal{B}\rangle -  \langle \bar{\epsilon} \rangle \\
   &\approx& \langle \mathcal{F}\rangle - \langle \mathcal{B}\rangle \quad \ \ \, \quad \qquad \text{if} \ \langle \mathcal{E}\rangle \gg \langle \bar{\epsilon} \rangle \ \text{(hydraulic control,   figure~\ref{fig:energetics_mean_fluxes}\emph{b-c})}\\
   &\approx&  (1-R_f)  \langle\mathcal{F}\rangle \\
      &\approx& 0.25 \,(1-R_f) \, \theta\, Ri_b^s \quad  \text{if} \  \langle \mathcal{F} \rangle \approx 0.25 \, \theta \, Ri_b^s \ \text{(figure~\ref{fig:energetics_mean_fluxes}\emph{a})}  \\
   &\approx&  0.037 \,   (1-R_f) \, \theta \qquad  \, \text{if} \  Ri_b^s \approx 0.15 \ \text{(Part 1, figure~2\emph{b})} \\
    &\approx&  0.035 \,  \theta \qquad \qquad \qquad  \, \text{if} \ R_f \approx 0.05 \ \text{(figure~\ref{fig:energetics_mean_fluxes}\emph{k})} \slabel{estim-E}
\end{subeqnarray}
where we recall that $\theta$ is in radians. Note that using the uncertain value of $\Gamma \approx 0.1$ in the last line \eqref{estim-E} (see figure~\ref{fig:energetics_mean_fluxes}\emph{j}) would give an almost identical expression $\langle \mathcal{E}\rangle \approx 0.034 \,  \theta$.

From \eqref{dKdt_0-3}, we propose the corresponding approximation of $\langle \chi \rangle$, in the limit of very turbulent flows with linear stratification where $\langle \mathcal{B}\rangle \approx \langle \mathcal{P}_\rho \rangle$ (figure~\ref{fig:energetics_mean_fluxes}\emph{l}):
\begin{equation}\label{estim-chi}
   \langle \chi \rangle \approx R_f \langle  \mathcal{F} \rangle  \approx 0.037 \, R_f \, \theta \approx 0.0019 \, \theta.
\end{equation}

We also note that, under all the above assumptions, our estimations \eqref{estim-E} and \eqref{estim-chi} yield the following ratio of scalar variance to kinetic energy dissipation:
\begin{equation}\label{estim-chi-over-E}
   \frac{\langle \chi \rangle}{\langle \mathcal{E}\rangle } \approx \frac{\langle \mathcal{B} \rangle}{\langle \mathcal{P}\rangle -\langle \mathcal{B}\rangle}  \approx \frac{R_f}{1-R_f}\approx \Gamma ,
\end{equation}
which gives values of $0.05$ or $0.1$, using our estimates of $R_f$ and $\Gamma$, respectively. This expression has the merit of linking $\Gamma,R_f$ with a natural measure of the irreversible `tax' levied by stratification on turbulence. The key question becomes: how do $\Gamma,R_f$ scale with the non-dimensional flow parameters? We tackle this parameterisation of mixing in \S~\ref{sec:param}.

Finally, we note that LPL19 explained the transitions between flow regimes by using the simple approximation $\langle \mathcal{E} \rangle \approx \langle \mathcal{P} \rangle \approx \langle \mathcal{F} \rangle \approx (h^2 \delta u ) \, \theta/8 \approx 0.04 \, \theta $ (where the factor $h^2 \delta u \approx 3$ converts their hydraulic non-dimensionalisation to our shear layer non-dimensionalisation). Their expression is in good agreement with \eqref{estim-E}. They argued that regime transitions are caused by thresholds in the normalised turbulent strain rate, which we write as
\begin{equation} \label{sijsij-theta-Re}
    \langle s'_{ij} s'_{ij} \rangle \equiv \frac{Re^s}{2} \langle \mathcal{E} \rangle \approx  0.02 \, (1-R_f) \, \theta  Re^s \propto \theta Re^s,
\end{equation}
assuming $R_f =$~const., highlighting the key role of the group of parameters $\theta  Re^s$.

The above data on mean energy reservoirs and fluxes confirm and extend LPL19's findings that flows with a similar product $\theta Re^s$ (but different individual values of $\theta$ and $Re^s$) behave similarly. Note that LPL19's hydraulic formulation used the product $\theta Re^h$, while our more accurate shear-layer formulation uses the product $\theta Re^s \propto \theta^{1.7} (Re^h)^{1.4}$. 

Our data are also consistent with the findings in Part 1 that quantitative turbulent fractions scale strongly with both $\theta$ and $Re^s$  (enstrophy fraction $\propto \theta^{2.7}(Re^s)^{2.8}$, and overturn fraction $\propto \theta^{3.2}(Re^s)^{1.8}$). Since the production of perturbation enstrophy by vortex stretching is given by $s'_{ij}\,\omega'_{i}\omega'_j$, there is in fact a direct relation between an increasingly large turbulent strain rate $s'_{ij} s'_{ij}$ (slaved to $\theta Re^s$) and increasingly extreme enstrophy events, and thus enstrophy fraction \citep{johnson_large-deviation_2016}. The relation to density overturns is more indirect; first because vorticity can be decomposed into a rotating and a shearing part \citep{tian_definitions_2018} (the rotating part being associated with overturns but not the shearing part), and second because overturns feed back into the enstrophy production through the baroclinic term.

\subsubsection{Kolmogorov and Batchelor length scales} \label{sec:energetics-Kolmogorov}

The estimation of the viscous dissipation of turbulent kinetic energy $\langle \mathcal{E} \rangle$ in \eqref{estim-E} allows us in turn to give a practical volume-averaged estimate of the Kolmogorov length scale $\ell_K$, marking the end of the inertial subrange for $K'$ and $K'_\rho$. Defined dimensionally as $(\nu^3/ \langle \mathcal{E}\rangle)^{1/4}$, its non-dimensional expression in shear layer units is
\begin{subeqnarray} \label{def-lK}
\ell_K &\equiv&  \langle \mathcal{E} \rangle^{-1/4}(Re^s)^{-3/4} \\ 
&\approx& 2  \, \theta^{-1/4} \, (Re^s)^{-3/4},
\end{subeqnarray}
assuming for simplicity that $1-R_f\approx 1$.

We also estimate the Batchelor length scale $\ell_B$, marking the end of the viscous convective sub-range for $K'_\rho$, as
\begin{equation}\label{def-lB}
\ell_B \equiv\ell_K Pr^{-1/2} \approx 0.1 \, \theta^{-1/4} \, (Re^s)^{-3/4}  \quad \text{for} \ Pr=700.
\end{equation}

These estimates give $\ell_K \approx 0.02$ and $\ell_B \approx 0.0007$ for T2 and T3. For these data sets, we thus only have suitable resolution in $x,z$ for the velocity field (since $dx = dz \approx 1.5\ell_K \approx 40 \ell_B$ and $dy \approx 5 \ell_K \approx 130 \ell_B$, see Part 1, Appendix~2). 

These estimates also suggest that while the magnitude of energy reservoirs and fluxes are strong functions of $\theta$, the Kolmogorov and Batchelor scales are stronger functions of $Re^s$ than of $\theta$. In particular, we note that in flows having identical `$\theta Re^s$ intensity', still have $\ell_K, \, \ell_B \propto (Re^s)^{-1/2}$, suggesting inherently different small-scale dynamics even at $\theta Re^s=$ const.

\subsection{Spatio-temporal profiles}\label{sec:energetics-spatiotemp}

\begin{figure}
\centering
\includegraphics[width=1.0\textwidth]{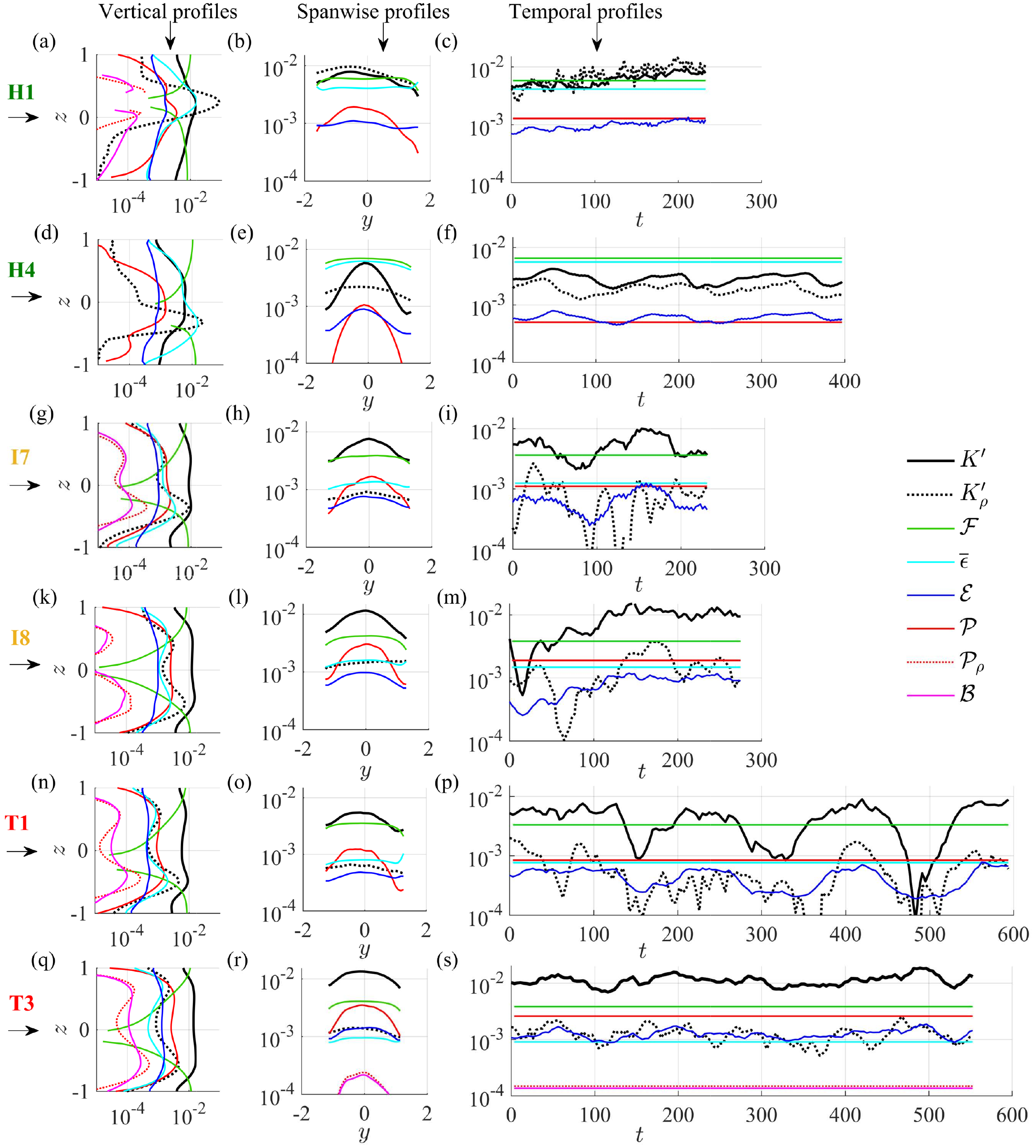}
    \caption{Profiles of turbulent energy reservoirs and fluxes in the vertical direction $z$ (left column); the spanwise direction $y$ (middle column); and time $t$ (right column) in six data sets: \emph{(a-c)} H1; \emph{(d-f)} H4; \emph{(g-i)} I7; \emph{(k-m)} I8; \emph{(n-p)} T1; \emph{(q-s)} T3. Axes limits and labels are identical in all panels of the left, middle, and right column, respectively. Note the semi-log scale in all panels. Data that is inferior to the lower axis limit are omitted (e.g. $\mathcal{F}$ partially $<0$ near $z=0$ in the left columns, and $\mathcal{P}_\rho, \mathcal{B}$ typically $<10^{-4}$ in the middle and right columns except in T3).  Also note that $\mathcal{F},\bar{\epsilon},\mathcal{P},\mathcal{P}_\rho,\mathcal{B}$ are by definition time-independent (right column).  }
    \label{fig:energetics_spatiotemporal}
\end{figure}

In figure~\ref{fig:energetics_spatiotemporal} we plot the vertical, spanwise, and temporal structure of the turbulent energy reservoirs $(K',K'_\rho)(\xx,t)$ and the volumetric fluxes $\mathcal{E}(\xx,t)$ and $(\mathcal{F},\bar{\epsilon},\mathcal{P},\mathcal{P}_\rho,\mathcal{B})(y,z)$. We show $z$ profiles in the left column (averaged in $x,y,t$ or $y$), the $y$ profiles in the middle column (averaged in $x,z,t$ or $z$),  and the $t$ profiles in the right column (averaged in $x,y,z$ or $y,z$). We only show six data sets whose 
energetics previously revealed interesting aspects representative of $\HH$ flows (H1 and H4, first and second rows), $\II$ flows (I7 and I8, third and fourth row) and $\TT$ flows (T1 and T3, fifth and sixth row, noting that T2 was omitted because it is similar to T3). The mean energy reservoirs $\bar{K},\bar{K}_\rho$ are omitted for clarity (but can be visualised by squaring $\bar{u},\bar{\rho}$ in Part 1, figure~3). Note that $\chi$ and $\Phi^{\bar{K}_\rho}$ are omitted too; the former because of its severe underestimation and low values (typically below the axes limits), and the latter as a consequence of our focus on kinetic energy budgets.

First, looking at the vertical profiles, $K'$ (in solid black) becomes nearly flat and symmetric over most of the shear layer as the flow becomes increasingly turbulent (panels~\emph{g,k,n,q}). The forcing $\mathcal{F}$ (in green) is always highest near the top and bottom edges of the shear layer (where $|\bar{u}|$ and $|\bar{\rho}|$ are highest) and vanishes in the middle (where it reaches slightly negative values, not shown on the log scale, where the $\bar{u}=0$ and $\bar{\rho}=0$ levels are offset). The turbulent dissipation $\mathcal{E}$ (in blue) closely matches the structure of $K'$ in all flows, albeit with approximately $1/10$ magnitude (giving an approximate turbulent dissipation time scale $K'/\mathcal{E} = O(10$~A.T.U.$)$). In T3 only, the turbulent dissipation exceeds the mean dissipation $\bar{\epsilon}$ (in cyan) throughout most of the shear layer (panel~\emph{q}). The mean dissipation $\bar{\epsilon}$ highlights the structure of the mean shear $\p_z \bar{u}$, typically higher on either side of the layer of mixed density, which matches more closely the structure of $K'_\rho$ than of $K'$. The scalar variance $K'_\rho$ (in dotted black) has a much sharper and sometimes asymmetric peak than $K'$, as seen in $\HH$ flows (symmetric Holmboe waves in panel~\emph{a}, asymmetric Holmboe waves in panel~\emph{d}) and some $\II$ flows (larger variance at the lower edge of the mixed layer in panel~\emph{g}). In $\II$ and $\TT$ flows, $K'_\rho$ tends to exhibit two peaks on either side of the mixed layer, due to overturning motions entraining fluid from the unmixed layers. In these flows the buoyancy flux $\mathcal{B}$ (in magenta) and production of scalar variance $\mathcal{P}_\rho$ (in dotted red) also tend to be nearly equal (as would be the case under linear stratification), and to closely match the structure of $K'_\rho$  (albeit with smaller magnitude, see panels~\emph{g,k,n,q}). Finally, in $\TT$ flows, the buoyancy flux $\mathcal{B}$ (in magenta) and the production of turbulent energy $\mathcal{P}$ (in red) have very similar profiles, corresponding to a uniform flux Richardson number $R_f(z)\approx 0.05$. This may be another hallmark of the self-organising equilibrium of stratified turbulent shear layers, related to the convergence of the gradient Richardson number to an equilibrium value $\approx 0.10-0.15$ as shown in Part 1.

Second, looking at the spanwise profiles, $K'$ nearly always has a sharper peak than the nearly-flat  $K'_\rho$ (panels~\emph{e,h,l,o,r}), a situation exactly opposite to that of their vertical profiles. The peak in $K'$ near $y=0$ is also much sharper than that of the mean flow $\bar{u}$ (see Part 1, figure~3), suggesting a peak in the ratio of turbulent-to-mean energy $K'/\bar{K}$ near $y=0$. This dichotomy between peaked \emph{vs}  flat spanwise profiles also extend to the turbulent fluxes $\mathcal{P},\mathcal{P}_\rho,\mathcal{B}$ \emph{vs}  the mean fluxes $\mathcal{F}$ and $\bar{\epsilon}$. Moreover, we know that outside the shear layer ($|y| > L_y$, $|z| > 1$) the turbulent fluxes decay to zero whereas the mean fluxes remains high. 

Third, in our interpretation of the $z$ and $y$ profiles, we recall that assuming a steady state and negligible boundary fluxes $\Phi^{\bar{K}},\Phi^{K'}$ should yield local (point-wise) equality of the following fluxes: $\langle \mathcal{F}\rangle_y \approx \langle \mathcal{P}\rangle_y + \langle \bar{\epsilon} \rangle_y$ and $\langle \mathcal{E}\rangle_y \approx \langle \mathcal{P}\rangle_y - \langle \mathcal{B} \rangle_y$ at all $z$ (and \emph{vice versa}, equality of $z$ averages at all $y$, as in \eqref{dKdt_0_prelim}).
In other words, these fluxes need to approximately balance everywhere both in $z$ and $y$ for $K'(\xx,t)$ to be steady (in term of curves: `green $=$ red $+$ cyan' and `blue  $=$ red $-$ magenta'). As we see in the left and middle columns, this is rarely the case in our data, presumably because of errors in our measurements of some turbulent quantities, and because of some non-negligible boundary fluxes $(\Phi^{\bar{K}}, \overline{\Phi^{K'}})(y,z) \neq 0$ due to (i) the slight non-periodicity of SID flows in $x$: $\overline{\p_{x}(uK)}$, $\overline{\p_{x}(u'K')}$ ;  (ii) the inevitable advective transport of $K'$ across our artificial `shear layer': $\overline{\p_{y}(v' K')+\p_{z}(w' K')}$; (iii) the unknown work of the mean and turbulent pressures in $x,y,z$.

Fourth, looking at the temporal profiles, the amplitude of the fluctuations in $K',K'_\rho,\mathcal{E}$ is small in H1, H4, T3 (panels~\emph{c,f,s}), and much larger in I7, I8, T1 (panels~\emph{i,m,p}). This is consistent with our nomenclature of the $\II$ regime as intermittently turbulent, and with our previous finding that T1 is actually closer to $\II$ flows than to T2 and T3, whose fluctuations are steadily large. Moreover, $K'$ and $K'_\rho$ are not generally correlated in $\II$ and $\TT$ flows, i.e. the intensity of velocity and density fluctuations do not temporally vary hand-in-hand (as might be incorrectly generalised from the time- and volume-averaged statement \eqref{estim-chi-over-E}). 
Finally, we recall that the temporal profiles of $K'$ and $K'_{\rho}$ should only reflect (with opposite correlation) the profiles of $\mathcal{E}$ and $\chi$, respectively, since all other fluxes plotted are (by definition) time-independent, and boundary fluxes are neglected. The negative correlation between $K'$ and $\mathcal{E}$ is, however, not always observed in our data (in fact, both appear almost positively correlated in most panels). These last two findings are not surprising, especially in light of our findings in Part 1 (figure 7) that turbulent fractions based on enstrophy or overturning can be largely uncorrelated, due to spatial heterogeneity of turbulent patches and the non-periodicity of our measurement volume along $x$.


\subsection{Spectra}\label{sec:energetics-spectra}

We now delve deeper into the flow energetics by investigating their spectra. We start with spectra of the turbulent kinetic energy and scalar variance along $x$, before focusing on individual velocity components and all variables $x,y,z,t$.  The limitations in our measurements of turbulent energetics, frequently hinted at in the above sections, will be discussed in light of spectral results in \S~\ref{sec:energetics-spectra-errors}.

\subsubsection{Spectra of $K',K'_\rho$ in $x$}

\begin{figure}
\centering
\includegraphics[width=1.02\textwidth]{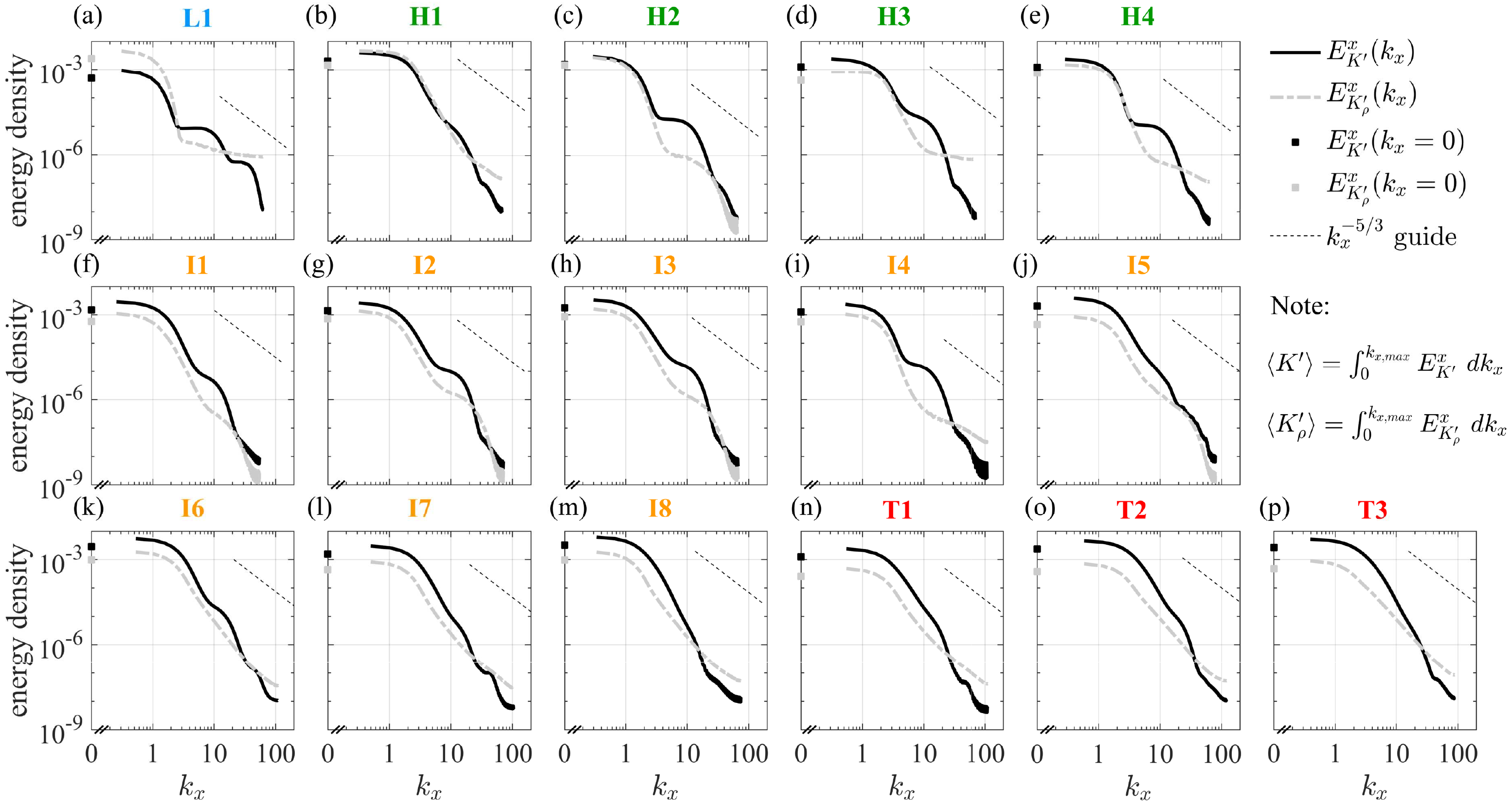}
    \caption{Spectral density along $x$ of the turbulent kinetic energy $E^x_{K'}$ and scalar variance $E_{K'_\rho}$ for all data sets, calculated with Welch's averaging method. The range of non-zero wavenumbers shown $[k_{x, \, min}, \, k_{x, \, max}] =[\pi/L_x, \pi/dx]$ varies slightly among data sets because of different domain lengths $2L_x$ and resolutions $dx$ (see Part 1, Appendix~2). Note the $k_x=0$ energy content (see text and Appendix~\ref{sec:Appendix-spectra} for details). Axes limits and labels are identical in all panels.}
    \label{fig:spectra_x_all}
\end{figure}

We define the spectral densities in $x$  of the mean turbulent kinetic energy $E^x_{K'}$ and scalar variance $E^x_{K'_\rho}$ such that
\begin{equation}\label{def-spectral-densities-in-x}
    \int_0^{k_{x, \, max}} E^x_{K'} \,  \d k_x = \langle K' \rangle , \qquad \int_0^{k_{x, \, max}} E^x_{K'_\rho} \,  \d k_x  =\langle K'_\rho \rangle.
\end{equation}
Their unambiguous definitions and the details of their practical computation from our discrete gridded data are given in Appendix~\ref{sec:Appendix-spectra}.
In the above, $k_{x, \, max} \equiv \pi/dx$ is the maximum (Nyquist) wavenumber that can be resolved in $x$. The (unusual) need to integrate from $k_x=0$ rather than from the minimum wavenumber $k_{x\, min} \equiv \pi/L_x$ comes from the fact that energy is contained in the mean ($k_x=0$), an inevitable consequence of the above definitions and of our definition of fluctuations around $x-t$ averages (more details in Appendix~\ref{sec:Appendix-spectra-energy-k0}).

In figure~\ref{fig:spectra_x_all}, we plot the densities $E^x_{K'}$ (black solid) and $E^x_{K'_\rho}$ (grey dashed) for all data sets. To correct for errors inherent to computing Fourier transforms of noisy and non-periodic data (over-estimating high-frequencies), here we plot estimations of these densities (i.e. periodograms) using Welch's averaging method. This standard method divides each original signal along $x$ into a series of overlapping segments, applies a window function to render them periodic, and returns the average square magnitude of their discrete Fourier transform (more details in Appendix~\ref{sec:Appendix-spectra-welch}).

Since we do not expect any turbulent signal in our laminar data set, the L1 spectra (panel~\emph{a}) are plotted as a `control', i.e. a baseline measure of inevitable artefacts due to the nature of our data and analysis. In panel~\emph{a}, $E^x_{K'}$ exhibits a distinct hump at intermediate wavenumbers ($k_x \approx 2-30$), correlated with a distinct hump or flattening of $E^x_{K'_\rho}$. This artefact is also found in varying degrees in most other data sets around $k_x \approx k_{x, \, max}/2$, affecting our most turbulent data to a lesser degree (panels~\emph{l-p}). 

Putting the above artefact aside, most $\HH,\II$ and $\TT$ spectra exhibit relatively similar shapes. The kinetic energy spectrum $E^x_{K'}$ is flat in the energy-containing range $k_x \lesssim 1$ (length scales $\gtrsim 6$), decaying as $k_x^{-\beta}$ in the inertial sub-range, with typical values around $\beta \approx 2.0-3.5$, and a slightly different power law decay near $k_{x,\,max}$.
The decay exponent $\beta$ is considerably larger than the classical Kolmogorov value $\beta=5/3\approx 1.7$ expected in isotropic turbulence \citep[\S~6.5]{pope_turbulent_2000}. This discrepancy in decay exponent may be due to the inherent low-pass filtering effect of PIV.

The scalar variance spectra $E^x_{K'_\rho}$ exhibit a  shape similar to $E^x_{K'}$, albeit with slightly smaller amplitude (as expected from the asymptotic 1/10 partition in figure~\ref{fig:energetics_mean_values}\emph{j}).  These spectra also have a smoother inertial sub-range decay extending all the way to $k_{x, \, max}$, at least in the most turbulent data (panels~\emph{k-p}), where the scaling  $k_x^{-\beta}$ is in better agreement with the expected value $\beta = 5/3$  \citep[\S~12.11]{kundu_fluid_2016}.

\subsubsection{Spectra of all components in $x,y,z,t$}

\begin{figure}
\centering
\includegraphics[width=1.03\textwidth]{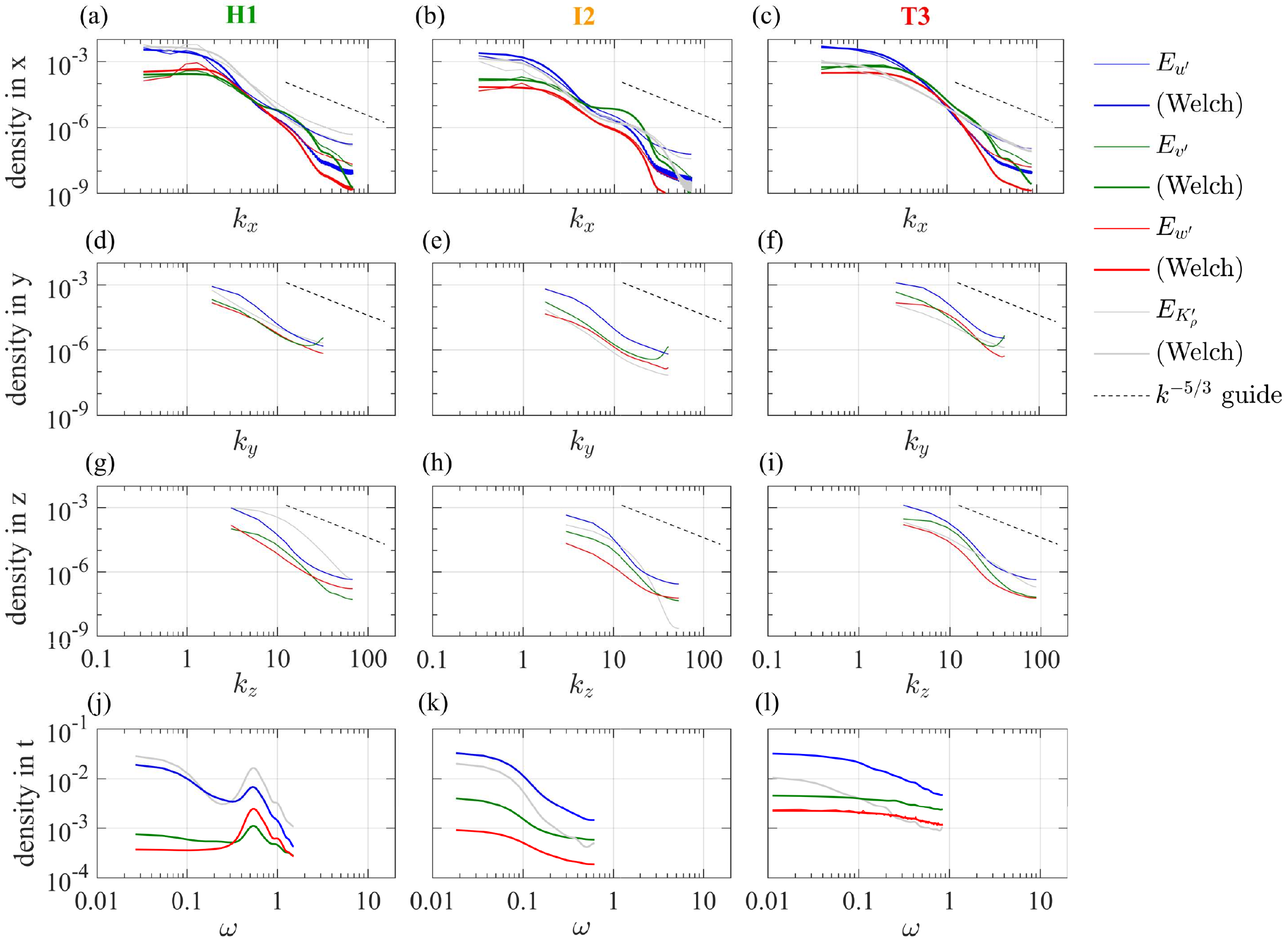}
    \caption{Spectral density of energy in individual velocity components $u'$ (blue), $v'$ (green), $w'$ (red), and of $K'_\rho$ (grey) in all directions $x$ \emph{(a-c)}, $y$ \emph{(d-f)}, $z$ \emph{(g-i)}, and  $t$ \emph{(j-l)} for three representative data sets H1 (left column), I2 (middle column), and T3 (right column). The mean energies $(1/2) \langle u'^2 \rangle$, $(1/2) \langle v'^2 \rangle$,  $(1/2) \langle w'^2 \rangle$, and $\langle K'_\rho \rangle$ are given by one-dimensional integration of any respective density, e.g. $(1/2) \langle u'^2 \rangle = \int_0^{k_x, \, max}  E^x_{u'} \d k_x = \int_0^{k_y, \, max} E^y_{u'} \d k_y = \int_0^{k_z, \, max} E^z_{u'} \d k_z = \int_0^{\omega, \, max} E^t_{u'} \d \omega$, etc. Mean values at $k_x,k_y,k_z,\omega=0$ are not shown for clarity. The spectral range depends on domain length and resolution $(k_x,k_y,k_z,\omega) \in [\pi/L_x, \pi/dx]\times  [\pi/L_y, \pi/dy] \times  [\pi, \pi/dz] \times  [2\pi/L_t, \pi/dt]$. Note the different axes scales between \emph{(a-i)} and \emph{(j-l)}. In $x$, we compare spectra obtained by the standard DFT (thin lines) and by Welch's method (thick lines). In $y,z$ we only show the former, and in $t$ we only show the latter (see text for more details).}
    \label{fig:spectra_xyzt}
\end{figure}

To gain further insight into the anisotropy of our flows and limitations of our energy spectra, we now extend our analysis to individual velocity components in all directions of space and time. We generalise the definition \eqref{def-spectral-densities-in-x} of $E^x_\phi$ as spectral densities in $x$ of $\phi= K', K'_\rho $ by defining $E^\xi_{\phi}$ where $\xi = x,y,z,t$ and $\phi=K', K'_\rho,u', v', w'$. The one-dimensional integrals of these spectral densities between their respective bounds of $0$ and $k_{x, \, max},k_{y, \, max},k_{z, \, max},\omega_{max}$ give respectively $\langle K'\rangle$,  $\langle K'_\rho \rangle$, $\langle u'^2/2 \rangle$, $\langle v'^2/2 \rangle$, $\langle w'^2/2 \rangle$, such that $E^\xi_{K'} = E^\xi_{u'}+E^\xi_{v'}+E^\xi_{w'}$ (for the full expressions see Appendix~\ref{sec:Appendix-spectra-discrete-def}). Note that the non-periodic and non-uniform gridded nature of our data prevents us from defining a single meaningful one-dimensional spectrum as often done in the isotropic turbulence literature (for more details see Appendix~\ref{sec:Appendix-spectra-3DFT}).

In figure~\ref{fig:spectra_xyzt} we plot these spectral densities in $x$ (top row), $y$ (second row), $z$ (third row), and $t$ (bottom row) for three representative data sets H1 (left column), I2 (middle column), and T3 (right column). To investigate the effects of the non-periodicity of our data on the energy spectra, we plot in $x$ the energy densities obtained using the standard discrete Fourier transform (DFT) periodogram (thin lines) and using Welch's estimated periodogram (thick lines, as in figure~\ref{fig:spectra_x_all}). Note that we only show the Welch in $t$ for conciseness, and we only show the DFT in $y$ and $z$ because the smaller numbers of data points in these directions render Welch's segmentation inappropriate (more details in Appendix~\ref{sec:Appendix-spectra-welch}).

We see in panels~\emph{a-c} that the standard DFT (thin lines) consistently overestimates high wavenumber content ($k_x\gtrsim 30$) compared to the Welch (thick lines), as expected from the fact that the latter is designed to minimise the effects of edge discontinuities in our data, incorrectly rendered as high-wavenumber energy by the standard DFT (called spectral leakage, or Gibbs phenomenon). Given these observations, we should remain critical in our interpretation of DFT spectra in $y,z$ (panels~\emph{d-i}), despite the fact that some of them exhibit an inertial sub-range decay closer to $k^{-5/3}$ in $y$ and $z$ than in $x$ (possibly due to spectral leakage countering the effects of PIV filtering). 

We further see that $u'$ usually has the most energy across all wavenumbers and frequencies, and $E_{u'}>E_{v'}>E_{w'}$. The lowest energy in $w'$ is consistent with the expectation that vertical motions are partially hindered by the stable stratification at $Ri_b^s>0$. The higher energy in $u'$ than in $v'$, particularly clear at very low streamwise wavenumbers $k_x\lesssim 1$, is partly due to our definition of fluctuations around $x-t$ averages and to the fact that the flow is not perfectly parallel (i.e. $u'$ can have a slight residual large-scale variance along $x$, as explained in Appendix~\ref{sec:Appendix-spectra-energy-k0}). 

The above observation that $E_{u'}>E_{v'}>E_{w'}$ has a few notable exceptions. 
First, we diagnose that the hump artefact in $E^x_{K'}$ observed in most panels of figure~\ref{fig:spectra_xyzt} appears primarily caused by $v'$ since  $E^x_{v'}>E^x_{u'}$ at medium and high $k_x$ (green lines in panels~\emph{a-c}), independently of the method (DFT or Welch). This artificial medium-scale structure in $v'(x)$  may come from the delicate stereo PIV calculation of $v$ (the component perpendicular to the laser sheet).  
Second, H1 exhibits $E_{v'}\approx E_{w'}$ across most wavenumbers  (panels~\emph{a,d,g}) and even $E_{v'}<E_{w'}$ in the frequency range $\omega \approx 0.3-1.5$ (panels~\emph{j}), which is consistent with the presence of Holmboe waves, known to generate vigorous vertical motions even in the presence of strong stratification (here $Ri_b^s = 0.567$).

The signature of Holmboe waves is indeed clear in the H1 temporal spectra at $\omega \approx 0.5$ (panel~\emph{j}), and also detectable in the longitudinal spectra $E^x_{w'}, E^x_{K'_\rho}$ around $k_x \approx 1$ (panel~\emph{a}, thin red and grey lines). These peaks suggest a typical phase speed $c\approx \omega/k_x \approx 0.5$ in agreement with observations in the spatio-temporal domain (not shown here). We also note in intermittent flow I2 a similar, albeit fainter, peak in all longitudinal spectra (panel~\emph{b}, thin lines), suggesting the faint presence of similar waves, in agreement with observations near the laminar/turbulent transitions (not shown here). Such spectral peaks are absent in the turbulent flow T3 (right column), suggesting dynamics across a broader range of spatio-temporal scales.


\subsection{Discussion and limitations} \label{sec:energetics-spectra-errors}


Based on the above insight from our energy spectra, we identify six key effects limiting the accuracy of our direct laboratory measurements of energy reservoirs and fluxes: (i) non-periodic and finite-length data; (ii) PIV and LIF filtering; (iii) resolution of the Kolmogorov and Batchelor length scales; (iv) volume reconstruction and spanwise distortion; (v) temporal resolution and aliasing; and  (vi) finite differentiation. We provide more details on each item in Appendix~\ref{sec:appendix-limitations-spectra}.

These limitations apply in particular to $\langle \mathcal{E}\rangle$ and $\langle \chi \rangle$, for which proxies were proposed in \S~\ref{sec:energetics-estimations-E-chi} (anticipating these limitations).

Indirect estimations in spectral space appear an attractive alternative to such direct estimations in physical space. A method can be conceived of as follows: the energy spectra of $E_{K'},E_{K'_\rho}$ are fitted to known theoretical `model' (ansatz) spectra, multiplied  by $k^2$ to yield the corresponding dissipation spectra, and integrated to obtain $\langle \mathcal{E} \rangle, \langle \chi \rangle$. Such spectra could include a $k^{-5/3}$ inertial sub-range scaling until $2\pi/\ell_K$ for $E_{K'},E_{K'_\rho}$, and a $k^{-1}$ viscous convective sub-range scaling until $2\pi/\ell_B$ for $E_{K'_\rho}$. However, this method has its own limitations. The inhomogeneity and anisotropy of our flows, key in the computation of $\p_{x_j} u'_i$ and $\p_{x_i} \rho'$, would require separate manipulation of the spectra of $u'^2, v'^2,w'^2,\rho'^2$ in each direction $k_x,k_y,k_z$, and \emph{a priori} knowledge of $\ell_K$ and $\ell_B$ (for which the estimations \eqref{def-lK}-\eqref{def-lB} could be used). Although scaling arguments and various \emph{ad hoc} anisotropy assumptions have been used (e.g. \citealp[\S~2]{hafeli_PIV_2014}), these remain speculative and would require further scrutiny. We are also mindful of the cautionary tale of \cite{okino_decaying_2019} who showed that high-$Pr$ density fluctuations can produce strong \emph{anisotropic} inertial-range velocity fluctuations down to $\ell_K$.

To make progress in this direction, the anisotropy of the velocity field is treated next in \S~\ref{sec:anisotropy}, while the parameterisation of turbulent energetics  is treated last in \S~\ref{sec:param}.


\section{Anisotropy}\label{sec:anisotropy}






 
Anisotropy is expected in SID flows due to the symmetry-breaking effects of the streamwise forcing, mean shear, vertical stratification (and, perhaps, the  boundary conditions of the apparatus).
In this section we investigate the large-scale anisotropy of the velocity field (controlling the production $\mathcal{P}$) in \S\S~\ref{sec:anisotropy-lumley}-\ref{sec:anisotropy-spatial-profiles}, followed by the small-scale anisotropy of the velocity gradients (controlling the dissipation $\mathcal{E}$) in \S~\ref{sec:anisotropy-diss}.

\subsection{Reynolds stresses and Lumley triangle}\label{sec:anisotropy-lumley}
 
We recall that the turbulent kinetic energy  $K' \equiv (1/2) \, \text{tr} \, \overline{\uu' \otimes \uu'}$ is the isotropic part of the Reynolds stress tensor  (half the trace of the one-point, one-time velocity cross-correlation tensor). By the Cauchy-Schwartz inequality, this diagonal part (isotropic `pressure') sets a bound on the magnitude of the off-diagonal part (deviatoric stresses): $K' \ge |\overline{u'v'}|, |\overline{u'w'}|,$ or $|\overline{v'w'}|$ \citep[eq. 5.109]{pope_turbulent_2000}. 
In idealised isotropic turbulence, all deviatoric  stresses are zero, thus there is no transfer between mean and turbulent kinetic energy, hence $\mathcal{P}=0$. By contrast, in shear-driven turbulence, this bound becomes more meaningful due to the crucial production of  $K'$ at rate $\mathcal{P}>0$ resulting from $\overline{u'w'} \neq 0$, i.e. from the net correlation of anisotropic eddies at large (energy-containing) scales. 

To quantify this anisotropy of the Reynolds stresses, we consider the widely-used normalised velocity anisotropy tensor $\mathsf{b}$, defined as the deviatoric part of the normalised Reynolds stress tensor $\overline{\uu' \otimes \uu'}/(2\bar{K'})$ with components 
\begin{equation}
\quad b_{ij}(y,z) \equiv  \frac{\overline{u'_i u'_j}}{\overline{u'_l u'_l}}- \frac{\delta_{ij}}{3},
\end{equation} 
as in \cite{pope_turbulent_2000} (\S~11.3.2). Since by definition $\text{tr} \, \mathsf{b} = b_{ii} =  0$ (first invariant), this tensor has only two independent invariants: $I\!I_b \equiv \text{tr} \, \mathsf{b}^2/2$ (second invariant) and $I\!I\!I_b  \equiv \det \mathsf{b} = \text{tr} \, \mathsf{b}^3/3$ (third invariant), which are more conveniently defined in normalised form as
\begin{equation}
\eta(y,z) \equiv \sqrt{\frac{I\!I_b}{3}}, \qquad \xi(y,z)\equiv \sqrt[3]{\frac{-I\!I\!I_b}{2}}.
\end{equation} 

\begin{figure}
\centering
\includegraphics[width=1.0\textwidth]{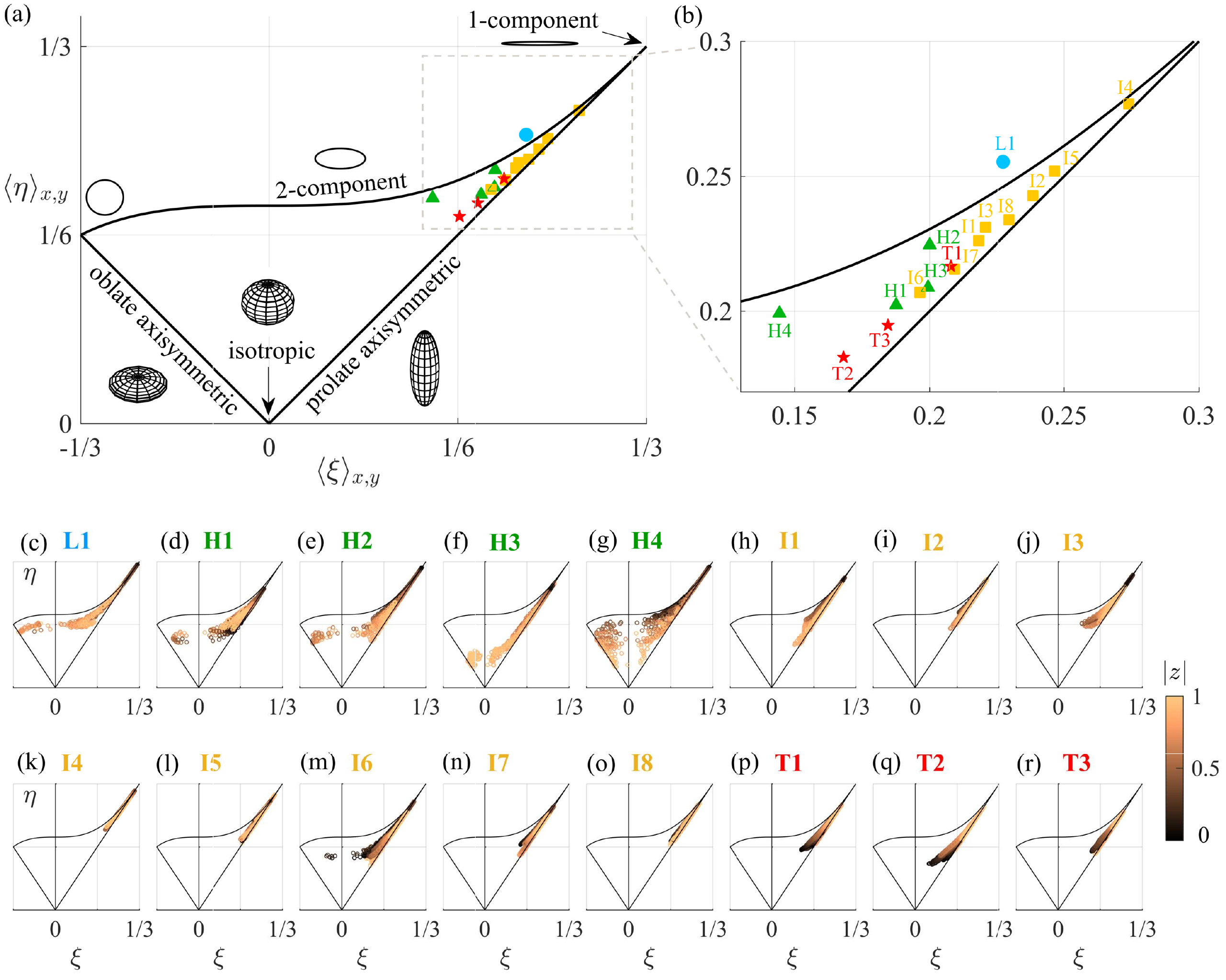}
    \caption{Degree and shape of Reynolds stress anisotropy in all 16 data sets. \emph{(a,b)} Mean values $\langle \xi\rangle_{y,z},\langle \eta\rangle_{y,z}$ (zoomed in detail in \emph{b}). The Lumley triangle is highlighted by thick lines, and the limiting cases of turbulence are shown schematically with principal axes coordinates. \emph{(c-r)}  All $n_y n_z$ data points of $(\xi,\eta)(y,z)$, coloured with the absolute vertical coordinate $|z|$ within the shear layer. }
    \label{fig:anisotropy_Lumley}
\end{figure}

The local state of anisotropy of a turbulent flow at any point $y,z$ can therefore be described by a point in the $\xi-\eta$ plane, lying inside the so-called \emph{Lumley triangle} \citep{lumley_computational_1978}, drawn with thick lines in figure~\ref{fig:anisotropy_Lumley}\emph{(a)}. The point $\xi=\eta=0$ corresponds to isotropic turbulence; the left (resp. right) straight edge $\eta=\mp \xi/2$ correspond to oblate (resp. prolate) axisymmetric turbulence, i.e. one principal eigenvalue being smaller  (resp. larger) than the other two; and the top curved edge $\eta=\sqrt{1/27+2\xi^3}$ corresponds to two-component turbulence (one principal eigenvalue being zero). In summary, the vertical axis $\eta$ quantifies the degree of anisotropy, while the horizontal axis $\xi$ quantifies its shape (oblate $\xi<0$ \emph{vs} prolate $\xi>0$).

In figure~\ref{fig:anisotropy_Lumley}\emph{(a-b)} we plot the mean $\langle \xi \rangle_{y,z}$ and $\langle \eta\rangle_{y,z}$ in all 16 data sets. First, we observe in panel~\emph{a} that all points are clustered in a narrow top-right region of strong prolate anisotropy, shown in greater detail in panel~\emph{b}. This is consistent with our prior spectral observation that the streamwise velocity perturbations dominate over the other two: $|u'|^2>|v'|^2,|w'|^2$.   Second, our `control' data set L1, being non-turbulent and therefore primarily affected by unphysical artefacts, distinguishes itself by being the only data set lying outside of Lumley's realisability triangle (though all $y,z$ points are by construction inside, the $y-z$ average does not have to be since this `curved triangle' is not convex). Third, asymmetric $\HH$ flows (H2, H4) lie closer to the two-component (top) limit, while symmetric $\HH$ flows (H1, H3) and most $\II$/$\TT$ flows lie closer to the prolate axisymmetric (right) limit. Fourth, almost all $\II$ flows exhibit stronger anisotropy than $\HH$ and $\TT$ flows, and lie closer to the one-component limit. This is a result of greater temporal variability (intermittency) in the streamwise component $u' \equiv u -  \langle u \rangle_{x,t}$, defined with respect to the streamwise and temporal average. We verified that removing intermittency effects by defining perturbations with respect to the streamwise average alone ($\uu -  \langle \uu \rangle_{x}$) did move $\II$ flows sligtly away from the one-component limit, but it did not change the qualitative picture of panels~\emph{a,b}. We also verified that removing streamwise variance effects ($\p_x \bar{u} \neq 0$) by defining perturbations with respect to the temporal average alone ($\uu -  \langle \uu \rangle_{t}$) changed the picture very little. In other words, $u'$ always dominates and anisotropy is not significantly biased by our definition of $u' \equiv u -  \langle u \rangle_{x,t}$ (used throughout Part 1 and Part 2).

In figure~\ref{fig:anisotropy_Lumley}\emph{(c-r)} we plot the underlying $n_y n_z$ data points within the triangle for each data set (of which panels~\emph{a-b} showed the centre of mass), to highlight the full range of anisotropy in $y$ and more particularly in $z$ (in colour). In all  flows, we observe large spatial variations around the mean, closely following the prolate axisymmetry limit (right edge), i.e. a state in which $v',w'$ have nearly (but not exactly) equal magnitude, while being dominated by $u'$, no matter the $y,z$ location.  This general trend is nuanced by the following subtleties. First, $\HH$ flows exhibit the greatest variations, and are unique in that they include pockets of \emph{oblate} anisotropy ($\xi<0$) for $|z|\approx 0.3-1$. Second, some $\II$/$\TT$ flows (I3, I6, T1, T2) have data points at $|z|\lesssim 0.3$ which deviate significantly away from the right edge (axisymmetry) and lie closer to the centre of the triangle. Third, the data points closest to the mid-point of the shear layer ($|z|\lesssim 0.2$, in black) tend to be the most anisotropic (largest $\eta$) in $\HH$/$\II$ flows but the least anisotropic in $\TT$ flows (smallest $\eta$).

These findings are qualitatively consistent with the unforced DNS of \cite{smyth_anisotropy_2000}, who observed  oblate axisymmetry during the initial growth of the Kelvin-Helmholtz instability and the turbulent transition, followed by prolate axisymmetry during the turbulent and decay phases (see their figure 6).

\begin{figure}
\centering
\includegraphics[width=0.95\textwidth]{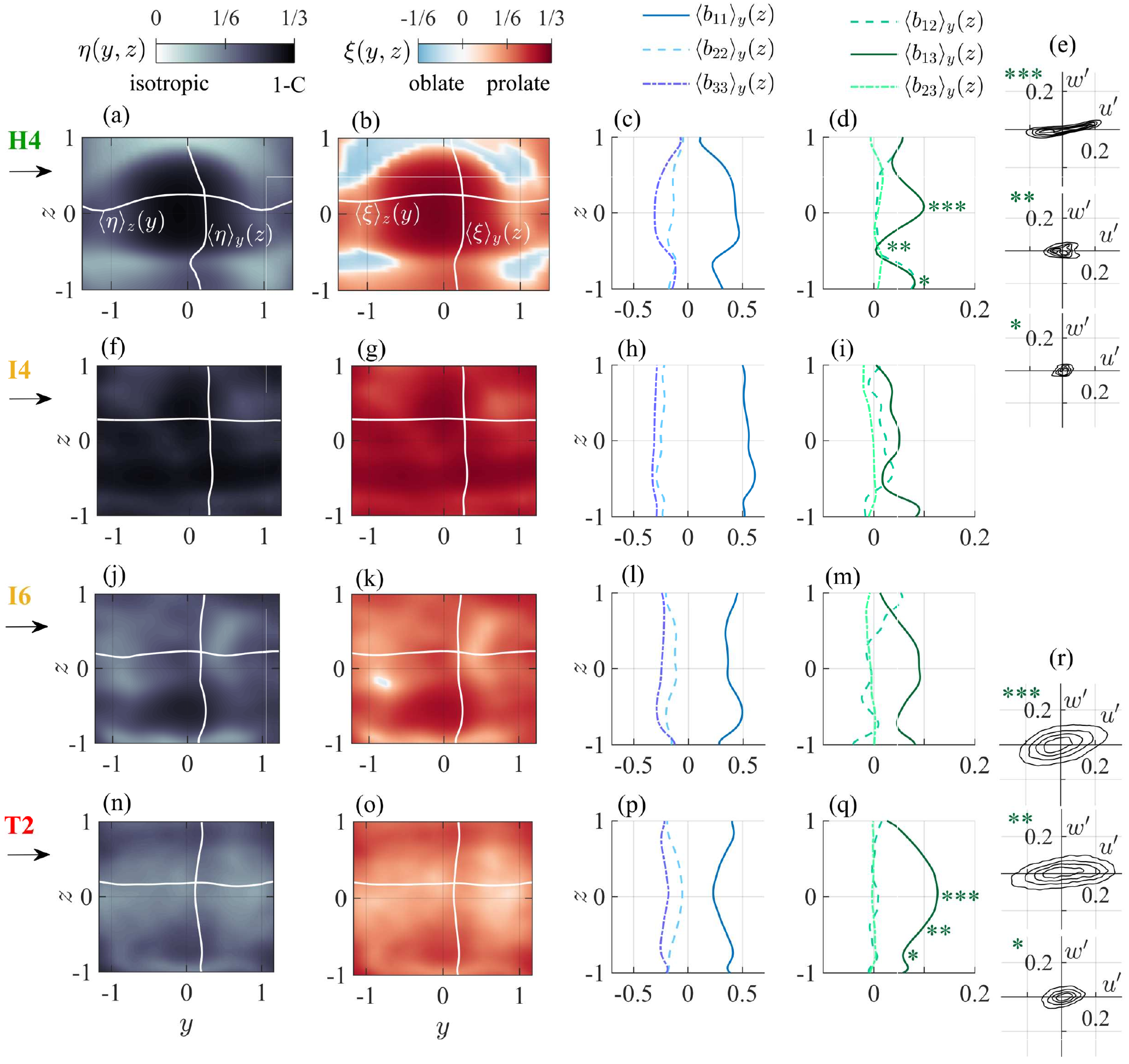}
    \caption{Spatial structure of the anisotropy tensor $\mathsf{b}$ of data sets H4, I4, I6, T2 (top to bottom row). Left to right column: second invariant $\eta(y,z)$, third invariant $\xi(y,z)$  (including the averages in each direction, superposed in white); vertical structure of the diagonal components $b_{11},b_{22},b_{33}$ averaged in $y$;  vertical structure of the off-diagonal components $b_{12},b_{13},b_{23}$ averaged in $y$. In \emph{(e,r)} we also show the p.d.f. of the $(u',w')$ clouds (rescaled histogram, here four equidistant contours at $20,40,60,80\,\%$) for H4 and T2  at three different vertical locations flagged by asterisks in \emph{(d,q)} (these are $z\in[-1, -0.9], [-0.55, -0.45], [0.05, 0.15]$ in H4, and $z\in[-0.9, -0.8], [-0.55, -0.45], [-0.05, 0.05]$ in T2, noting that in both we restrict the region to $|y|\le 0.5$ to show a stronger signal).  }
    \label{fig:anisotropy_profiles}
\end{figure}


\subsection{Spatial profiles}\label{sec:anisotropy-spatial-profiles}
 
To delve deeper into these tantalising observations, we plot in  figure~\ref{fig:anisotropy_profiles} the spatial structure of $\eta(y,z)$, $\xi(y,z)$, and the vertical structure of the six individual components of the (symmetric) tensor $\langle (b_{ij}) \rangle_y(z)$ for the four representative data sets H4, I4, I6 and T2. In the contrasting cases of H4 and T2, we also plot the underlying probability density function (p.d.f.) of the  $(u',w')$ data at three distinct vertical locations.

First, starting with the $y-z$ structures (colour plots in the left two columns), we find that the region of weak oblate anisotropy in H4 (light grey in panel \emph{a} and light blue in panel \emph{b})  lies at the periphery of a  core of strong prolate anistropy (this subtle structure is lost in the $y$ and $z$ averages superimposed in white). We explain this oblate pocket by the particular structure of confined Holmboe waves described in \cite{lefauve_structure_2018} in this same H4 flow, and in particular by the large values of $v'$ and its odd symmetry about the $y=0$ axis, responsible for the divergence and convergence of streamlines in $z=$~const. planes around the upward-pointing crests of the density interface (see their figure 8\emph{(k,l)} and point (v) in \S~6.1.2). By contrast, $\II$/$\TT$ flows have a more uniform structure, ranging from strong prolate anisotropy in I4 (panels \emph{f,g}) to weaker and less prolate anisotropy in I6  (panels \emph{j,k}) and T2 (panels \emph{n,o}) especially in the most turbulent region $|z|\lesssim 0.5$.

Second, moving on to the diagonal components (third column), we confirm our above claims that $b_{11}=\overline{u'^2}/(2\bar{K'})-1/3$ (and therefore $u'$) dominates in all flows, where it approaches its upper bound of $2/3$ in the most anistropic $z$ locations (darkest colours in the first two columns), while the complementary  $b_{22},b_{33}$ approach their lower bound of $-1/3$.  Furthermore,  $b_{22}>b_{33}$ in all flows (and therefore $v'$  dominates over $w'$), a natural consequence of stratification inhibiting vertical motions. 

Third, moving on to the off-diagonal components (fourth column), we recall that they are all bounded above and below by $\pm 1/2$ by the Cauchy-Schwarz inequality.  We find that $b_{13}\equiv\overline{u'w'}/(2\bar{K'})$ dominates over $b_{12},b_{23}$ almost everywhere in all flows, being always positive and even reaching $\langle b_{13}\rangle_y(z=0)= 0.125$ in T2 (panel~\emph{q}), i.e. no less than 25\,\% of its upper bound. This proves that while $w'$ contributes less than $v'$ to the reservoir $K'$, it contributes much more than $v'$ to the production of $K'$ because of its much greater correlation with $u'$, recalling that the logarithmic production rate $\p_t (\ln \sqrt{K'}) = \p_t K'/(2K')$ is $\mathcal{P}/(2K')\equiv -b_{12}\p_y \bar{u}-b_{13}\p_z \bar{u}$. The broad peak of $b_{13}$ in the vigourous flows I6 and T2 (panels~\emph{m,q}) is absent in the less vigorous flow I4 and in the Holmboe flow H4 (panels~\emph{d,i}), the latter having instead two narrower peaks at $z\approx -1$ and $z\approx 0$.

Fourth, these contrasting  $\langle b_{13}\rangle_y(z)$ profiles in H4 and T2 can be understood by their respective p.d.f.s in the $u'-w'$ plane (fifth column). In the Holmboe flow (panel~\emph{e}) the $z\approx -1$ peak  (denoted by $\ast$) is due to a compact but fairly tilted distribution towards the first and third quadrant ($u'w'>0$), while the $z\approx -0.5$ trough ($\ast\ast$) is due to a broader but more up-down symmetric distribution, and the $z\approx 0$ peak ($\ast\!\ast\!\ast$) is due to a yet broader, but thinner and more tilted distribution. In the turbulent flow (panel~\emph{r}) the increase of $\langle b_{13}\rangle_y(z)$ with decreasing $|z|$ is due to a broader distribution (compare $\ast$ and $\ast\ast$) followed by an increased tilt (compare $\ast\ast$ and $\ast\!\ast\!\ast$). 

Fifth, it is possible to improve the quantification of this tilt of $(u',w')$ distributions by investigating the orientation of the principal axes of $\mathsf{b}$, given by its eigenvectors assembled in a matrix $\bm{V}$ such that $\bm{b}\equiv(b_{ij})=\bm{V}\bm{\Lambda} \bm{V}^{-1}$ (where $\bm{\Lambda}$ is the diagonal matrix of principal eigenvalues). To avoid the intricate analysis of three Euler angles describing the three-dimensional rotation matrix $\bm{V}$, we  take advantage of the fact that $|b_{12}|,|b_{23}|\ll|b_{13}|$ in T2 to simplify the analysis to a single angle $\beta$ describing the two-dimensional rotation $\tilde{\bm{V}}$ (around the $v'$ axis) of the reduced $\tilde{\bm{b}}\equiv [b_{11}\ b_{13} ; b_{13}, b_{33}]$. This angle $\beta(y,z)$ locally quantifies the tilt between  the $u'$ axis and the major principal axis, and we therefore expect $0<\beta \ll 90^\circ$ based on  panel~\emph{r}.  Although we do not plot it for conciseness,  the profile of $\langle \beta\rangle_y(z)$ follows almost exactly that of $\langle b_{13}\rangle_y(z)$, with a minimum value of $5^\circ$ at $|z|=1$ and a maximum value of $16^\circ$ at $|z|=0$, in excellent agreement with the qualitative insights derived from panel~\emph{r}.


\subsection{Velocity gradients and dissipation surrogates}\label{sec:anisotropy-diss}

We now investigate the anisotropy of velocity gradients, controlling the rate of turbulent dissipation, which is by definition the sum of 12 squared gradient terms belonging to three key groups:
\begin{equation}
\begin{aligned}\label{eq:gradient_terms}
     \langle \mathcal{E} \rangle  \equiv \dfrac{2}{Re^s} \langle s_{ij}'s_{ij}' \rangle = \dfrac{2}{Re^s}  &\Big\langle \underbrace{(\p_x u')^2  + (\p_y v')^2 +   (\p_z w')^2}_\text{longitudinal} 
     \\
     &+  \underbrace{(\p_y u')^2  + (\p_z u')^2 +   (\p_x v')^2 + (\p_z v')^2  + (\p_x w')^2 +   (\p_y w')^2}_\text{transverse} 
     \\
     & + \underbrace{\p_y u'\p_x v' + \p_z u'\p_x w' + \p_z v'\p_y w'}_\text{asymmetric}  \Big\rangle.
\end{aligned}
\end{equation}
In idealised homogeneous isotropic turbulence, all terms belonging to the same group (longitudinal, transverse, or asymmetric) are equal. Using the continuity equation $\p_{x_i} u'_i=0$, it can further be shown that any transverse term is twice as large as any longitudinal term (e.g. $\langle (\p_y u')^2 \rangle = 2\langle (\p_x u')^2 \rangle$, etc) while any asymmetric term is negative and only half as large (e.g. $\langle \p_y u' \p_x v' \rangle = (-1/2)\langle (\p_x u)^2 \rangle$, etc) \citep{almakie_energy_2012}. Plugging in these relations into \eqref{eq:gradient_terms} allows to estimate $\langle \mathcal{E} \rangle$ under the assumption of isotropy using only one term (instead of 12), as follows
\begin{equation}\label{eq:dissipation_surrogates}
    \langle \mathcal{E} \rangle \approx \ \begin{cases} 
   \ \ \frac{15}{Re^s}\langle (\p_{x_i} u_i')^2\rangle  & \text{(longitudinal surrogate)} \\[6pt]
   \ \ \frac{15}{2Re^s}\langle (\p_{x_i} u_j')^2\rangle,  \quad \ i\neq j& \text{(transverse surrogate)} \\[6pt]
    -\frac{30}{Re^s}\langle \p_{x_i} u_j'\p_{x_j} u_i' \rangle, \ \ i\neq j & \text{(asymmetric surrogate)}  
    \end{cases}
\end{equation}
where, importantly, we do not sum over repeated indices here. These simple one-dimensional and one-component surrogates have been used for decades in laboratory and field measurements due to the difficulty of measuring more than one or two terms (although there exists more sophisticated multi-component models that relax isotropy, and e.g. assume axisymmetry instead). Our data sets provide us with the complete set of twelve terms in all directions $(x,y,z,t)$ and thus allow us to test the validity of these (time- and volume-averaged) surrogates and, thus, of the underlying assumption of small-scale isotropy.

To do so, we arrange the above 12 candidates into the following `surrogate dissipation matrix' $\langle \varepsilon_{ij}\rangle$, and define its relative estimation error $\langle \tilde{\varepsilon}_{ij}\rangle$ as
\begin{equation}\label{eq:matrix_surrogates}
 \langle \varepsilon_{ij} \rangle \equiv \frac{15}{Re^s} 
 \begin{bmatrix}
 \langle (\p_x u')^2 \rangle & \frac{1}{2}\langle (\p_y u')^2\rangle & \frac{1}{2}\langle (\p_z u')^2\rangle \\[9pt]
 \frac{1}{2}\langle (\p_x v')^2\rangle & \langle (\p_y v')^2 \rangle & \frac{1}{2}\langle (\p_z v')^2\rangle \\[9pt]
\frac{1}{2}\langle (\p_x w')^2\rangle & \frac{1}{2}\langle (\p_y w')^2\rangle & \langle (\p_z w')^2 \rangle \\[9pt]
-2\langle \p_y u'\p_x v' \rangle & -2\langle \p_z u'\p_x w' \rangle  & -2\langle \p_z v'\p_y w' \rangle
\end{bmatrix},
 \quad \langle \tilde{\varepsilon}_{ij} \rangle \equiv \frac{\langle \varepsilon_{ij} \rangle  - \langle \mathcal{E} \rangle }{\langle \mathcal{E} \rangle } ,
\end{equation}
where the top $3\times 3$ block contains the three longitudinal terms (diagonal) and the six transverse terms (off-diagonal), as in \cite{portwood_asymptotic_2019}. The fourth row contains the three asymmetric terms; these are rarely used in applications because they are impractical, but we include them nonetheless to obtain a complete picture of small-scale anisotropy.

\begin{figure}
\centering
\includegraphics[width=0.75\textwidth]{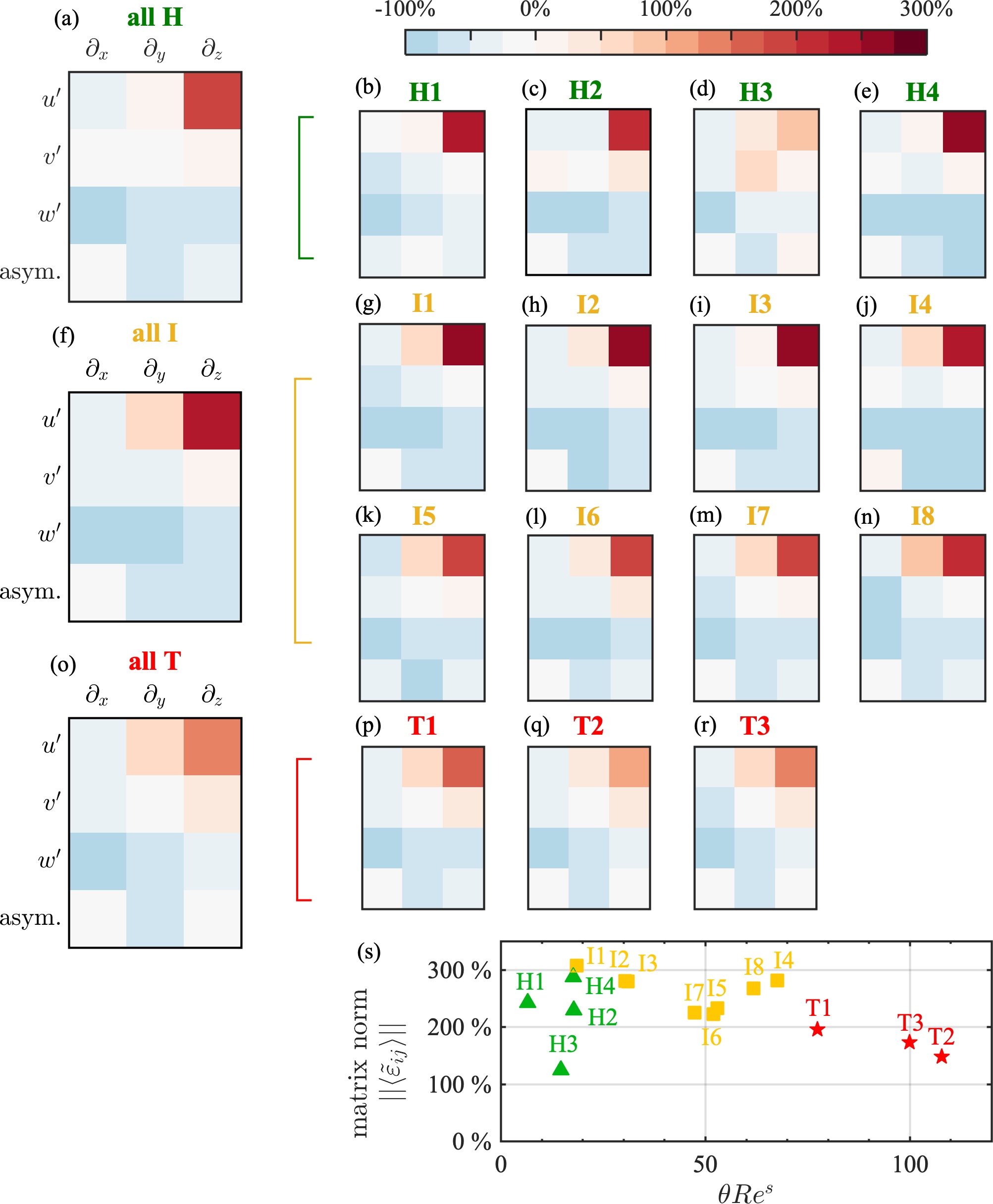}
    \caption{Anisotropy of the 12 density gradients in  \eqref{eq:gradient_terms}, measured by the error made by using them as surrogates for $\langle \mathcal{E} \rangle$ based on the assumption of isotropy (as in \eqref{eq:dissipation_surrogates}) in all H, I, and T data sets. Colours show the value of each entry of the $4 \times 3$ matrix of relative error $\langle \tilde{\varepsilon}_{ij} \rangle$ defined in \eqref{eq:matrix_surrogates}. All H, I, T data sets are shown, together with their mean across each regime in the left-most panels \emph{(a,f,o)}. Blue indicates an underestimation, red indicates an overestimation, and darker shades indicate poorer estimation, i.e. stronger anisotropy. \emph{(s)} Matrix norm quantifying dissipation anisotropy \emph{vs} $\theta Re^s$ (isotropy corresponds to $||\langle \tilde{\varepsilon}_{ij} \rangle|| = 0\,\%$. }
    \label{fig:anisotropy_dissipation}
\end{figure}

In figure~\ref{fig:anisotropy_dissipation} we visualise  the relative error matrix $\langle \tilde{\varepsilon}_{ij} \rangle$ in all  $\HH$, $\II$ and $\TT$ flows, together with their `intra-regime mean' on the left. Each entry in each $4\times 3$ matrix is coloured according to its departure away from isotropy; a negative value means that the surrogate is an underestimation (in blue, bounded below by $-1$), while a positive value means that the surrogate is an overestimation (in red, not bounded above but always $<3$ here).

Focusing first on the individual data sets (right part of the figure) and considering the global pattern of each matrix, we find strong similarities between all flows, and more specifically, between all flows within a same regime ($\HH, \II$, or $\TT$), with perhaps only one exception in H3 (panel~\emph{d}). Importantly, terms that are clearly positive (resp. negative) are robustly so across most flows. This implies that intra-regime means of each matrix entry do not artificially cancel out values of opposite signs (which would incorrectly imply isotropy) and, therefore, that these means give a meaningful representative picture of each regime.

Focusing then on these robust means (panels~\emph{a,f,o}), we also find similarities between them. First, $(\p_z u')^2$ (top right term) is consistently overwhelming, and overestimates $\langle \mathcal{E}\rangle$ by as much as $200\,\%$ ($\HH$ flows), $230\,\%$ ($\II$ flows), or $140\%$ ($\TT$ flows). This can be attributed to the influence of the mean shear $(\p_z u')^2$. In $\II$/$\TT$ flows, $(\p_y u')^2$ and $(\p_z v')^2$ also tend to consistently overestimate $\langle \mathcal{E}\rangle$, whereas they are reliable in $\HH$ flows. Second $(\p_x w')^2$ consistently underestimates by $80-90\,\%$, while all four other terms involving $w'$ gradients (bottom right $2\times 2$ block) consistently underestimate by $20-70\,\%$. This can be attributed to the stable mean stratification, hindering vertical motion. Third, all $x$ gradients (first column of each matrix) are generally weak. This can be attributed to the elongation of flow structures along $x$ by the mean shear. Fourth, the best estimates (lightest shade) varies slightly from regime to regime, but three terms stand out as consistently reliable: $(\p_y v')^2$, $(\p_z v')^2$, and $\p_y u' \p_x v'$  (having $<35\,\%$ relative error everywhere, sometimes much less). 

Finally, we plot the Euclidian (Frobenius) norm of each matrix  $||\langle \tilde{\varepsilon}_{ij} \rangle|| \equiv (\sum_{i,j} \langle \tilde{\varepsilon}_{ij} \rangle^2)^{1/2}$ (panel~\emph{s}) against the product of parameters $\theta Re^s$, identified in \eqref{sijsij-theta-Re} as controlling the dissipation (norm of the strain rate tensor). The general trend is a decrease of small-scale anisotropy with increasing $\theta Re^s$ (stronger turbulence), from typical values of $200-300\,\%$ in $\HH$-$\II$ flows (except H3) to values below $200\,\%$ in $\TT$ flows. This trend suggests that even stronger turbulence ($\theta Re^s \gg 100$) would continue to approach greater isotropy, as indeed observed by \cite{itsweire_turbulence_1993},  \cite{smyth_anisotropy_2000} (see their figure 14), \cite{hebert_relationship_2006}, and most recently by \cite{portwood_asymptotic_2019} (see their figure 2) with increasing `dynamic range', quantified by the buoyancy Reynolds number $Re_b$.  We define $Re_b$ in the next section, explain its relation to $\theta Re^s$, and introduce other ratios of kinematic and dynamic scales to tackle  parameterisations.

\section{Parameterisations} \label{sec:param}

In this section we study the parameterisation of turbulent fluxes using simple flow quantities such as mean gradients or scalar parameters. After providing the background and definitions of various measures of mixing and parameterisation approaches in \S~\ref{sec:param-background}, we assess these parameterisations in \S~\ref{sec:param-summary1}-\ref{sec:param-summary3} with an in-depth analysis of data sets I6-T3 to seek `asymptotic' scaling laws valid in  strongly turbulent flows.

\subsection{Background: measures of mixing} \label{sec:param-background}

\subsubsection{Direct measures: eddy diffusivities}  \label{sec:param-direct}

Stratified turbulent mixing is usually modelled in large-scale circulation models by a single parameter, the  eddy (or turbulent) diffusivity for the stratifying agent (heat or salt) $\kappa_T$, and for the momentum $\nu_T$. This turbulence closure scheme relies on the simple turbulent flux / mean gradient relations (see \cite{pope_turbulent_2000} Chap. 10)
\begin{equation} \label{def-nuT,kappaT}
\frac{\kappa_T}{Re^s}  \equiv   \frac{-\overline{w'\rho'}}{\p_z\bar{\rho}}    \equiv   \frac{\mathcal{B}}{\Nbb^2},  \qquad \frac{\nu_T}{Re^s} \equiv   \frac{-\overline{u'w'}}{\p_z\bar{u}} \approx \frac{\mathcal{P}}{\Sbb^2}.
\end{equation}
The approximation in $\nu_T$ reflects the fact that production is dominated by the vertical shear $|\overline{u'w'}\p_z\bar{u}| \gg |\overline{u'v'}\p_y\bar{u}|$ in our flows.  Importantly, the $Re^s$ factor comes from the fact that we choose to define both eddy diffusivities as non-dimensional ratios relative to the molecular value for momentum $\nu$, rather than relative to the (default and implicit) inertial scale $(\Delta U \delta u H h)/16$ (recall Part 1, \S~3.2-3.3). Other authors legitimately choose to define $\kappa_T$ relative to the molecular value for the scalar $\kappa$, which then gives $\kappa_T/(Re^s Pr) \equiv \mathcal{B}/\Nbb^2$. Also recall (Part 1, \S~5) the definitions of the square buoyancy frequency $\Nbb^2 \equiv -Ri_b^s \, \p_z\bar{\rho}$ and square shear frequency $\Sbb^2 \equiv (\p_z \bar{u})^2$ based on the mean flow (the double overline avoids confusion with $\overline{\p_z\rho}$ and $(\overline{\p_zu})^2$ which are different quantities not discussed here). The gradient Richardson number based on the mean flow was defined as $\Rigbb \equiv \Nbb^2/\Sbb^2$.

Despite being used as the `direct' (or `ultimate') measures of mixing in most practical models, eddy diffusivities are necessarily simplistic descriptions of the process of stratified turbulent mixing. They have been criticised for their apparent inability to address the complex underlying energetics, in particular to disentangle the partition between irreversible mixing and reversible stirring in $\mathcal{B}$ \citep{salehipour_diapycnal_2015}. However, upon inspection of the budget equation \eqref{dKdt}, we find that under linear stratification ($\mathcal{B}\approx \mathcal{P}_\rho$) and neglecting boundary fluxes ($\Phi^{K'_\rho}\approx 0$), the buoyancy flux appears to be in `lock step' with the irreversible dissipation of scalar variance $\mathcal{B}\approx \chi$, which, again under linear stratification, is equivalent to the dissipation of perturbation available potential energy, i.e. irreversible mixing \citep{caulfield_open_2020}. This led some authors to argue that defining $\kappa_T$ using $\chi/\Nbb^2$ was generally more appropriate than using $\mathcal{B}/\Nbb^2$, an approach known as the `Osborn-Cox method' after \cite{osborn_oceanic_1972} (see \cite{salehipour_diapycnal_2015,gregg_mixing_2018,taylor_testing_2019} for more details). Following this line of thought, some authors define the flux coefficient $\Gamma$ as $\chi / \mathcal{E}$, which agrees with our approximation \eqref{estim-chi-over-E}. However, unlike DNS data, our experimental data  do not allow us to access $\chi$ directly with good accuracy, which is why we pursue an indirect approach, discussed next.

\subsubsection{Indirect measures: flux coefficients, mixing lengths} \label{sec:param-indirect}

Starting with the original definition \eqref{def-nuT,kappaT}, we attempt to relate the elusive $\mathcal{B}$ to the more tangible $\mathcal{E}$ (the `turbulence intensity'). This approach proposes equivalent definitions for $\kappa_T,\nu_T$ using our previous definitions of $\Gamma, R_f, \Rigbb$ and of a new turbulent Prandtl number  $Pr_T$:
\begin{equation} \label{def-nuT,kappaT-osborn}
\kappa_T \equiv Re^s  \frac{\Gamma\langle\mathcal{E}\rangle}{\langle \Nbb^2 \rangle} \equiv \Gamma  Re_b, \qquad Pr_T \equiv  \frac{\nu_T}{\kappa_T} \equiv \frac{\langle\mathcal{P}\rangle}{\langle\mathcal{B}\rangle} \frac{\langle\Nbb^2 \rangle}{\langle\Sbb^2\rangle} \approx \frac{\langle \Rigbb \rangle}{R_f}  \approx \frac{1+\Gamma}{\Gamma}\, \langle\Rigbb\rangle,
\end{equation}
where the buoyancy Reynolds number $Re_b \equiv Re^s \langle\mathcal{E}\rangle / \langle \Nbb^2 \rangle$ is a measure of the `turbulence intensity' that we will return to in \S~\ref{sec:param-measures-length}. The first approximation in $Pr_T$ comes from $\langle \Nbb^2 \rangle / \langle \Sbb^2 \rangle \approx \langle \Nbb^2/\Sbb^2 \rangle$, and the second approximation comes from the approximate link between $\Gamma$ and $R_f$ in  \eqref{eq-link-gamma-Rf}, valid under the simplified balance \eqref{dKdt_0-1} of \cite{osborn_estimates_1980}.

Eddy diffusivites can also be expressed using the Prandtl mixing length model, which posits that the turbulent fluxes depend quadratically on the mean gradients:
\begin{equation} \label{def-mixing-lengths}
L_\rho^2 \equiv   \frac{-\overline{w'\rho'}}{|\p_z\bar{u}|\p_z\bar{\rho}}    \equiv   \frac{\mathcal{B}}{\Sbb \Nbb^2},  \qquad L_m^2 \equiv   \frac{-\overline{u'w'}}{|\p_z\bar{u}|\p_z\bar{u}} \approx \frac{\mathcal{P}}{\Sbb^3}, \qquad \frac{L_\rho^2}{L_m^2} = Pr_T,
\end{equation}
and therefore that $\kappa_T = L_{\rho}^2 \Sbb$, $\nu_T=L_m^2 \Sbb$ where  $L_\rho, L_m$ are the non-dimensional `mixing lengths' for density and momentum, respectively. They can be interpreted as the typical distance travelled by a fluid parcel before its density or momentum becomes mixed with its surroundings (analogous to the mean free path of a molecule in the kinetic theory of gases). The stratified shear flow experiments of \cite{odier_fluid_2009,odier_understanding_2012,znaien_experimental_2009} showed that $L_\rho,L_m$ were approximately uniform in $z$ (instead of $\kappa_T,\nu_T$), i.e. that the quadratic flux-gradient relationships \eqref{def-mixing-lengths} were better approximations than the linear flux-gradient relationships \eqref{def-nuT,kappaT}.

Nevertheless, putting this aside for now and assuming the validity of the widely-used eddy diffusivity model \eqref{def-nuT,kappaT}, the key challenge of parameterising $\kappa_T$  (and its related $\nu_T$) using \eqref{def-nuT,kappaT-osborn} becomes equivalent to parameterising the dependence of the indirect (or `proximate') parameter $\Gamma$ (or its related $R_f$) on a few key non-dimensional parameters best characterising the flow, an approach known as the `Osborn method' after \cite{osborn_estimates_1980} (see  \cite{salehipour_diapycnal_2015,gregg_mixing_2018,taylor_testing_2019} for more details). To achieve this, different dynamical balances have been proposed, based on the ratios of relevant length scales or time scales which we discuss next.

\subsubsection{Parameters based on length scales and time scales ratios} \label{sec:param-measures-length}

Further to our definitions in \S~\ref{sec:energetics-Kolmogorov} of the microscopic Kolmogorov length scale $\ell_K$ (see \eqref{def-lK}) and Batchelor length scale $\ell_B$ (see \eqref{def-lB}), we  now define the Ozmidov length scale $\ell_O$ and the Corrsin  length scale $\ell_C$, which represent the smallest scales at which the distorting influences of background stratification and shear, respectively, are felt \citep{smyth_length_2000}. Their non-dimensional expressions in shear layer units are
\begin{equation}\label{def-lO-lC}
\ell_O \equiv \Big( \frac{\langle \mathcal{E}\rangle}{\langle \Nbb^3\rangle}\Big)^{1/2} \equiv \frac{\langle \mathcal{E} \rangle^{1/2}}{ (Ri_b^s)^{3/4} \langle  |\p_z \bar{\rho}|^{3/4}\rangle}, \qquad \ell_C \equiv \Big( \frac{\langle \mathcal{E}\rangle}{\langle \Sbb^3\rangle}\Big)^{1/2} \equiv \frac{\langle \mathcal{E} \rangle^{1/2}}{  \langle  |\p_z \bar{u}|^{3/2}\rangle} 
\end{equation}
Note the subtle fact that the $y-z$ averaging (integration) is made after raising the power in the denominator, contrary to the numerator. This choice is often ambiguous in the literature, and in the following we average  $\Nbb,\Sbb$ sometimes before, and sometimes after raising the power, for notational convenience. However, this (common) abuse of notation is justifiable in the more strongly turbulent flows I6-T3 in which $\Nbb, \Sbb \approx $ const. (thus the power and integration operators commute with good accuracy). In these flows we have the following separation of scales
\begin{equation}\label{lO-lC-scaling}
    \frac{\ell_O}{\ell_C} \approx \frac{ \langle  |\p_z \bar{u}|^{3/2}\rangle}{\langle  |\p_z \bar{\rho}|^{3/4}\rangle}  (Ri_b^s)^{-3/4}  \rightarrow  (Ri_b^s)^{-3/4} \rightarrow 5.
\end{equation}
using $|\p_z\bar{u}|\approx |\p_z\bar{\rho}| \rightarrow 1$ and $Ri_b^s \rightarrow 0.15$. In other words, there exists a moderate range of eddy sizes that are significantly more influenced by shear than by stratification, i.e. the turbulence is slightly dominated by shear.

The separation between the Ozmidov and the Kolmogorov scales is usually quantified by the buoyancy Reynolds number $Re_b$ (first mentioned in \eqref{def-nuT,kappaT-osborn}):
\begin{subeqnarray}\label{def-Reb}
Re_b  \equiv  Re^s  \frac{\langle \mathcal{E} \rangle}{\langle \Nbb^2\rangle } = \Big(\frac{\ell_O}{\ell_K}\Big)^{4/3} &=& Re^s (Ri_b^s)^{-1}\frac{\langle \mathcal{E} \rangle}{  \langle |\p_z \bar{\rho}|\rangle}   \slabel{Reb_est0}  \\
&\rightarrow & 0.2 \, \theta Re^s \quad  \slabel{Reb_est1} \\
 &\approx & 10-20 \quad \ \text{for} \ \ \theta Re^s = 50-110 \slabel{Reb_est2} 
\end{subeqnarray}
The turbulent estimate \eqref{Reb_est1}  assumes: (i) $Ri_b^s\rightarrow 0.15$ (Part 1, figure 2\emph{b}), (ii) $\langle \mathcal{E}\rangle \rightarrow 0.035 \theta$ in \eqref{estim-E},  and (iii) $\langle |\p_z \bar{\rho}|\rangle\rightarrow 1$, the latter being verified to better than $5\,\%$ in I6-T3. This expression is slightly different from that of \cite{lefauve_regime_2019} who proposed $Re_b \rightarrow 0.12 \, \theta Re^h$ (see their equations (6.9) and (6.10)), using the hydraulic Reynolds number (instead of the shear-layer Reynolds number), and averaging data across the whole duct cross-section (instead of the `core' shear layer only). Our estimate yields \eqref{Reb_est2} in  I6-T3, in which $\theta Re^s = 50-110$ (see figure~\ref{fig:energetics_mean_fluxes}\emph{(c)}. We conclude from this comfortable separation of scales that there exists a significant range of eddy sizes that are too small to be significantly affected by stratification but too large to be dominated by viscous dissipation, which is a requirement for the existence of stratified turbulent dynamics.

The indirect measure of mixing $\Gamma$ has often been assumed constant $\approx 0.2$ by physical oceanographers (corresponding to the upper bound set by \cite{osborn_estimates_1980}, as mentioned in \S~\ref{sec:energetics-mean-fluxes}). The DNSs of \cite{shih_parameterization_2005} suggested that this constant value was indeed accurate in `transitional' turbulence ($Re_b\approx 7-100$), but that $\Gamma \propto Re_b^{-1/2}$ in `energetic' turbulence ($Re_b> 100$); a scaling that has been much debated and reinterpreted since.

It is now widely acknowledged that the challenge of isolating key non-dimensional parameter(s) controlling turbulent mixing was due to the pervasive tendency for these parameters to be correlated in often-unsuspected, flow-specific, and potentially misleading ways. As an example, \cite{maffioli_mixing_2016} and \cite{garanaik_inference_2019} recently argued that $\Gamma$ should not be a function of the (ambiguous)  parameter $Re_b$, but of a (more fundamental) turbulent Froude number $Fr$ instead. This Froude number is defined as the ratio of the turbulent kinetic energy dissipation frequency $\mathcal{E}/K'$ to the buoyancy frequency (or the ratio of the buoyancy time scale to the dissipation time scale):
\begin{subeqnarray}\label{def-Fr}
    Fr \equiv \frac{\langle \mathcal{E} \rangle}{ \langle K' \rangle \langle \Nbb \rangle} &\equiv& (Ri_b^s)^{-1/2} \langle |\p_z\rho|^{-1/2} \rangle  (\langle K' \rangle/\langle \mathcal{E} \rangle )^{-1}   \slabel{def-Fr-1} \\
    &\approx& 0.3, \slabel{def-Fr-2}
\end{subeqnarray}
using $Ri_b^s\rightarrow 0.15$,  $\langle |\p_z \bar{\rho}|\rangle\rightarrow 1$, and
the approximate energy dissipation time scale $K' / \mathcal{E} \approx 10$  observed in  \S~\ref{sec:energetics-spatiotemp}.  Their scaling analyses and triply-periodic, spectrally-forced DNSs suggest that  $\Gamma \approx 0.5 \propto Fr^0$  in strongly-stratified flows ($Fr\ll 1$); that $\Gamma \propto Fr^{-1}$ in  moderately-stratified flows ($Fr \approx 1$); and that $\Gamma \propto Fr^{-2}$ in weakly-stratified flows ($Fr\gg 1$). Note that their argument relies on a definition of $\Gamma$ using the ratio of irreversible components $\chi /  \mathcal{E}$, which is only consistent with our definition $\mathcal{B} /  \mathcal{E}$ under conditions of `lock step' between $\mathcal{B}$ and $\chi$ explained in \S~\ref{sec:param-direct} (asymptotically satisfied at large $\theta Re^s$). 

\begin{figure}
\centering
\includegraphics[width=0.97\textwidth]{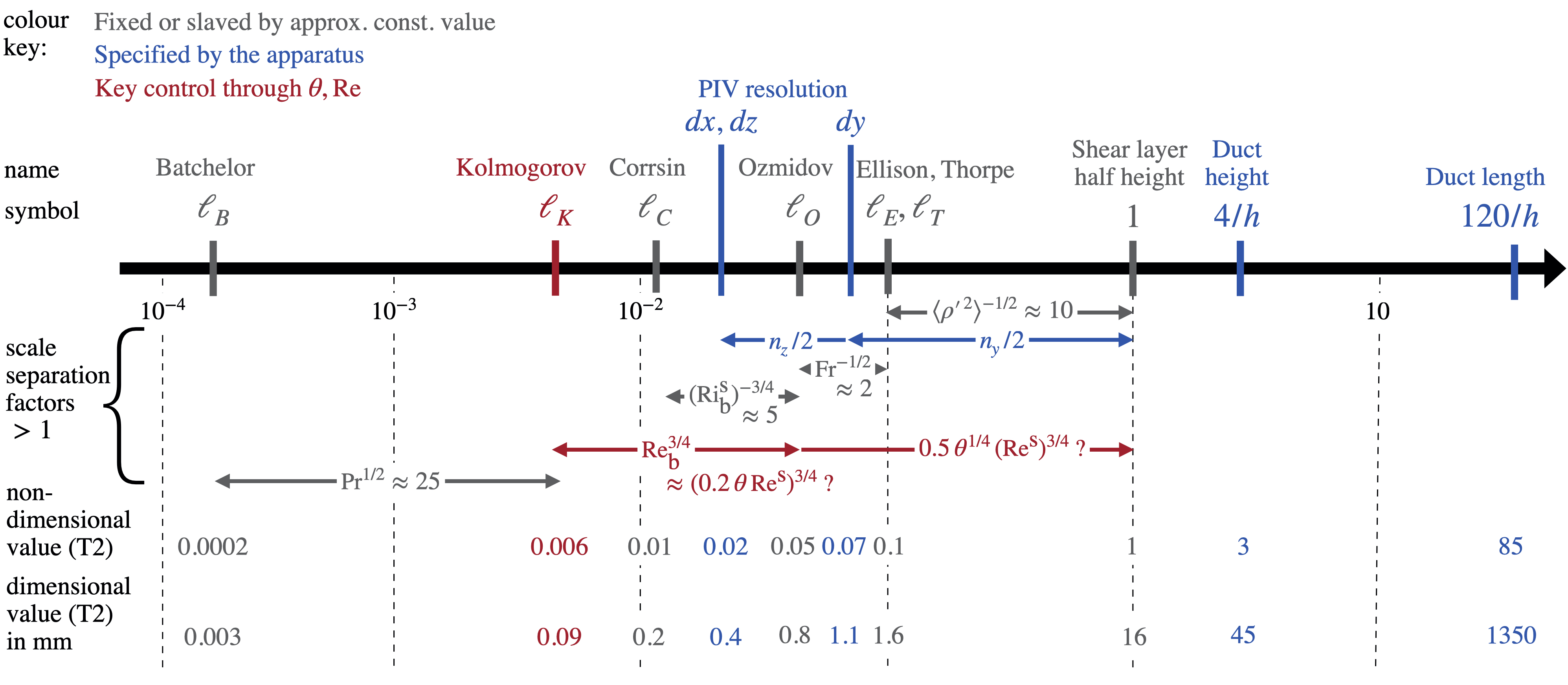}
    \caption{Key length scales in SID turbulence. Relative positions and scale separation factors are based on the simple volume-averaged estimates of  \S~\ref{sec:param-measures-length} in the asymptotic turbulent regime ($\theta Re^s \gg 100$).  Note the logarithmic axis.  Non-dimensional values (in shear layer variables) and dimensional values (in mm) are for data set T2. The Kolmogorov scale $\ell_K$ (in red) has two slightly incompatible scalings \eqref{def-lK} and \eqref{def-Reb}, flagged by `?' (the numerical values use the less trustworthy \eqref{def-Reb} for consistency with the other scales). Ideally the PIV resolution would approach $\ell_K$, and the LIF resolution (here assumed equal to the PIV) would approach $\ell_B$.}
    \label{fig:length-scales-schematics}
\end{figure}

This turbulent Froude number is connected to a further key  scale, the Ellison scale
\begin{equation} \label{def-lE}
    \ell_E  \equiv \frac{\langle \rho'^2 \rangle^{1/2} }{\langle |\p_z \rho|\rangle} \rightarrow 0.07-0.12,
\end{equation}
using $|\p_z \rho|\rightarrow 1$, and $\langle \rho'^2 \rangle \approx 0.005-0.015$ for I6-T3 as observed in figure~\ref{fig:energetics_mean_values}\emph{(e)}. It measures the typical vertical distance travelled by fluid parcels to achieve an stable equilibrium density profile through adiabatic sorting. It is closely related to the Thorpe scale $\ell_T$, defined directly on any  instantaneous vertical density profile as the root-mean-square of these sorting displacements  \citep{mater_relevance_2013}. \cite{garanaik_inference_2019} argued that $Fr \approx (\ell_O/\ell_E)^2$ when $Fr\ll 1$; that $Fr \approx L_O/L_E$ when $Fr \approx 1$; and that $Fr \approx (L_O/L_E)^{2/3}$ when $Fr\gg 1$. Our estimate \eqref{def-Fr-2} suggests that I6-T3 are relatively strongly stratified, hence that 
\begin{equation} \label{lO-lE-scaling}
   \frac{\ell_E}{\ell_O} \approx Fr^{-1/2} \rightarrow 2,
\end{equation}
i.e. that the separation between the Ellison and Ozmidov scales is very modest.

Figure~\ref{fig:length-scales-schematics} summarises the above estimates by showing the relative position of these length scales expected in the asymptotic turbulent regime. In addition to the general separation factors between different scales, we also give the corresponding specific values for each scale in data set T2 (non-dimensional value in shear layer unit and dimensional value in mm). In dark grey we highlight scales that are fixed (the reference shear layer half height $1$) or slaved by approximately constant parameters ($\ell_E$ in \eqref{def-lE}, $\ell_O$ in \eqref{lO-lE-scaling}, $\ell_C$ in \eqref{lO-lC-scaling}, and $\ell_B$ in \eqref{def-lB}). In blue we highlight scales that are specified by the apparatus and thus subject to change by the experimenter (the duct height $4/h$ and length $120/h$, as well as the PIV vector resolution, where $h,dx,dy,dz,n_x,n_y,n_z$ were given in Part 1, Table~1). In maroon we highlight the only scale, $\ell_K$, which is directly controlled through the two key variable flow parameters $\theta, Re^s$ ($\ell_K$ can be visualised as a `slider', unconstrained by other scales). Its definition \eqref{def-lK} and the use of our turbulent estimate  \eqref{estim-E} provided a (likely correct) scale separation factor of $\ell_K^{-1} \approx 0.5\, \theta^{1/4} (Re^s)^{3/4}$ (with respect to the shear layer scale $1$). However, the scaling arguments in this section yield a slightly incompatible  $\ell_K^{-1} \approx Re_b^{3/4} Fr^{-1/2} \langle \rho'^{2}\rangle^{-1/2} \approx 20 \, Re_b^{3/4} \approx 6\, \theta^{3/4} (Re^s)^{3/4}$ (combining three factors to reach the shear layer scale $1$). We believe this (likely incorrect) scaling in $\theta^{3/4}$ can be explained by a weak, neglected  dependence of $Fr$ and  $\langle \rho'^{2}\rangle$ on $\theta$ (i.e. $\ell_O$ and $\ell_E$ are not exactly constant).






\subsubsection{Objectives}

In the next three sections \S\S~\ref{sec:param-summary1}-\ref{sec:param-summary3} we will analyse data sets I6-T3, beyond the simple volume averages used above, with two specific objectives.

First, dimensional analysis suggests that all measures of mixing, from the direct eddy diffusivities, to the indirect flux coefficients, to the key dynamical parameters $Re_b, Fr$ should generally be functions of our five non-dimensional parameters $(\theta,Re^s,Ri^s_b,R,Pr)$.  Since we have a fixed $Pr=700$, and $Ri_b^s \approx 0.15, R \approx 2$, we will only probe the dependence on $\theta$ and $Re^s$.
 
Second, a slight abuse of notation in the above must be acknowledged: $\kappa_T,\nu_T$ in \eqref{def-nuT,kappaT-osborn} used time- and volume (bracket) averages and are scalar quantities uniform in space (like $\Gamma, Rf$ in \eqref{def-Gamma-Rf}, and $Re_b, Fr$), whereas $\kappa_T,\nu_T$  in \eqref{def-nuT,kappaT} used only $x-t$ (bar) averages and were functions of $y,z$. 
Our second  objective in the next sections will therefore be to  use all data points in $y-z$ to examine the hitherto implicit relevance of using uniform values for $\kappa_T,\nu_T,L_m,L_\rho,Pr_T,\Gamma,R_f,Re_b,Fr$, and, thus, of the implicitly-assumed linear relationships between their respective numerators and denominators. 

We tackle eddy diffusivities, mixing lengths and the turbulent Prandtl number in \S~\ref{sec:param-summary1},  the flux coefficient,  flux Richardson number  (as well as the $\mathcal{B}/\mathcal{P}_\rho$ ratio) in \S~\ref{sec:param-summary2}, and finally the buoyancy Reynolds number and turbulent Froude number in \S~\ref{sec:param-summary3}.

\subsection{Eddy diffusivities, mixing lengths, turbulent Prandtl number} \label{sec:param-summary1}

\begin{figure}
\centering
\includegraphics[width=0.98\textwidth]{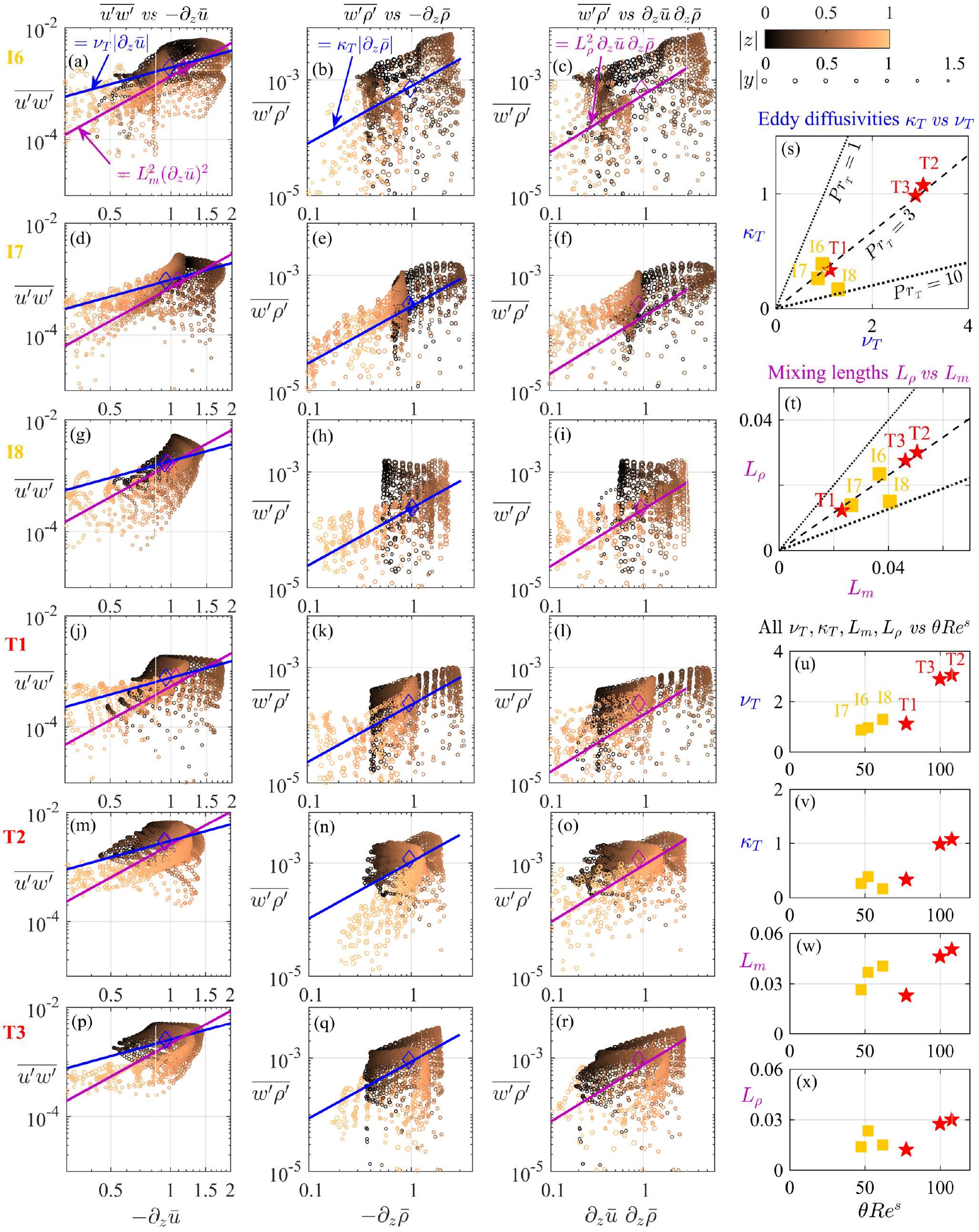}
    \caption{Eddy diffusivities and mixing lengths in data sets I6-T3 (top to bottom row). Clouds of $n_y n_z$ points of $\overline{u'w'} \equiv \mathcal{P}/\Sbb$ \emph{vs} $-\p_z \bar{u} \equiv \Sbb$ (first column);  $\overline{w'\rho'} \equiv \mathcal{B}/Ri_b^s$ \emph{vs} $-\p_z \bar{\rho} \equiv N^2/Ri_b^s$ (second column);   $\overline{w'\rho'} \equiv \mathcal{B}/Ri_b^s$ \emph{vs} $\p_z \bar{u} \, \p_z \bar{\rho}\equiv \Sbb \Nbb^2/Ri_b^s$ (third column). Note the log-log axes, and symbol colour and size respectively indicating the $|z|$ and $|y|$ location. Linear and quadratic least-squares fits provide the eddy diffusivities (in blue, after multiplying by $Re^s$ as in \eqref{def-nuT,kappaT}) and mixing lengths (in purple). Diamonds show the volume-averaged values  of the flux \emph{vs} gradient (blue) or square gradient (purple).  Right column:  $\nu_T,\kappa_T,L_m,L_\rho$ values \emph{(s-t)} against one another, giving the ratio $Pr_T=\kappa_T/\nu_T = L_\rho^2/L_m^2$ ($Pr_T= 1, 3, 10$ shown);   \emph{(u-x)} against the input parameters $\theta Re^s$. Values obtained from the fit are indistinguishable from those obtained from the diamonds.}
    \label{fig:parameterisation_1}
\end{figure}

In figure~\ref{fig:parameterisation_1} we test the flux-gradient relations \eqref{def-nuT,kappaT},~\eqref{def-mixing-lengths} with the full clouds of $n_y n_z$ data points (left three columns). Linear fits with enforced zero intercept are plotted in blue, and provide the eddy diffusivities $\nu_T, \kappa_T$, while quadratic fits with enforced zero intercept are shown in purple, and provide the mixing lengths $L_m,L_\rho$. These `fit' values are then plotted in the rightmost column.

First, focusing on the left three columns, we find that the clouds generally have a wide spread, making the fits fairly poor (also note the log-log axes). Despite this spread, the fits capture a clear monotonic tendency, particularly visible  in the upper boundary of each cloud (high flux values) which are indeed bounded by an approximately linear or quadratic flux relation. The symbol colours, indicating $|z|$, reveal that these high flux values tend to occur close to the mid-point of the shear layer ($|z|\approx 0$, dark colour), though less so in the buoyancy flux (second column). The symbol sizes, inversely proportional to $|y|$, do not reveal any clear correlation between flux-gradient behaviour and spanwise location, other than the fact that $|y|$ contributes to the spread of the clouds. Although the coefficients of determination are generally very low ($r^2<0.2$), the constant eddy diffusivity model (linear fit) does slightly better than the constant mixing length model (quadratic fit) overall. 
However, this is not very significant because uniform eddy diffusivities and uniform mixing lengths  actually become compatible in our asymptotic case of uniform shear $\Sbb \approx 1$, since by definition $(\kappa_T/Re^s) \equiv L_\rho^2 \, \Sbb$ and $(\nu_T/Re^s) \equiv L_m^2 \, \Sbb$. In other words, our range of $\Sbb \equiv -\p_z \bar{u} \approx 0.5-1.5$ (see left column) is not wide enough to convincingly argue in favour of either model.

Second, the diamond symbols show the volume average of the flux -- the numerator -- against the volume-average of the gradient  (in blue) or square gradient (in purple) -- the denominator. As expected from our above comment that uniform eddy diffusivities and mixing lengths are compatible,  blue and purple diamonds lie close to one another, near the horizontal values of $1 \approx \langle \Sbb\rangle \approx \langle \Sbb^2 \rangle \approx \langle \Nbb\rangle  \approx \langle \Sbb \Nbb\rangle$. Moreover, most diamonds sit very close to the fits (lines) of their respective colour in the left two columns, but consistently above them in the third column. This proves that the definitions of $\nu_T,\kappa_T,L_m$ by volume averages would produce good approximations of the fit of the underlying distribution (i.e. the fit goes through the centre of mass of the cloud), whereas the definition of $L_\rho$ by volume averages would produce an overestimation.

Third, moving on to the rightmost column, we find good correlations $\kappa_T \propto \nu_T$ (panel~\emph{s}) and $L_\rho \propto L_m$ (panel~\emph{t}), corresponding to a constant turbulent Prandtl number $Pr_T \approx 3$ (dashed line), except in I8 which has $Pr_T\approx 7$. This is entirely consistent with the approximation in \eqref{def-nuT,kappaT-osborn} and our previously quoted asymptotic values of $\Rigbb \approx 0.15$ (Part 1, \S~5) and $R_f\approx 0.05$ (\S~\ref{sec:energetics-mean-fluxes}) giving $Pr_T\approx 3$. This value is comfortably above 1, despite the tendency to self-similarity of the mean velocity and density profiles observed in $\TT$ flows ($\langle \bar{u} \rangle_y(z) \approx \langle \bar{\rho} \rangle_y(z)$, see Part 1, figure 3). This value is however consistent with the DNSs of \cite{salehipour_diapycnal_2015} (see their figure 10, at higher $Re^s$ but similar $Ri_b^s$)  who found $Pr_T \approx 3$ at $Re_b\approx 5-15$. Actual values for the diffusivities range from $\nu_T\approx 1$ in I6-T1 to $\nu_T\approx 3$ in T2-T3,  a substantial but not overwhelming increase with respect to the molecular value for momentum $\nu$. The corresponding range $\kappa_T Pr \approx 700/3-700$ is, by contrast, an overwhelming increase with respect to the molecular value for density $\kappa$, i.e. a high `eddy P\'eclet number'. Mixing lengths $L_m,L_\rho$ are of the order of the Kolmogorov length $\ell_K$ (see estimate in \eqref{sec:energetics-Kolmogorov}) and of the resolution of our measurements in $x,z$ (see Part 1, table 3). Finally, all quantities typically increase  monotonically with $\theta Re^s$ (panels~\emph{u-x}), though T1 is an outlier that appears  less energetic than suggested by its $\theta Re^s$ value. We conclude that $\nu_T,\kappa_T$ appear linear or superlinear in $\theta$ and $Re^s$.

\subsection{Flux coefficient, flux Richardson number, and $\mathcal{B}/\mathcal{P}_\rho$} \label{sec:param-summary2}

\begin{figure}
\centering
\includegraphics[width=0.98\textwidth]{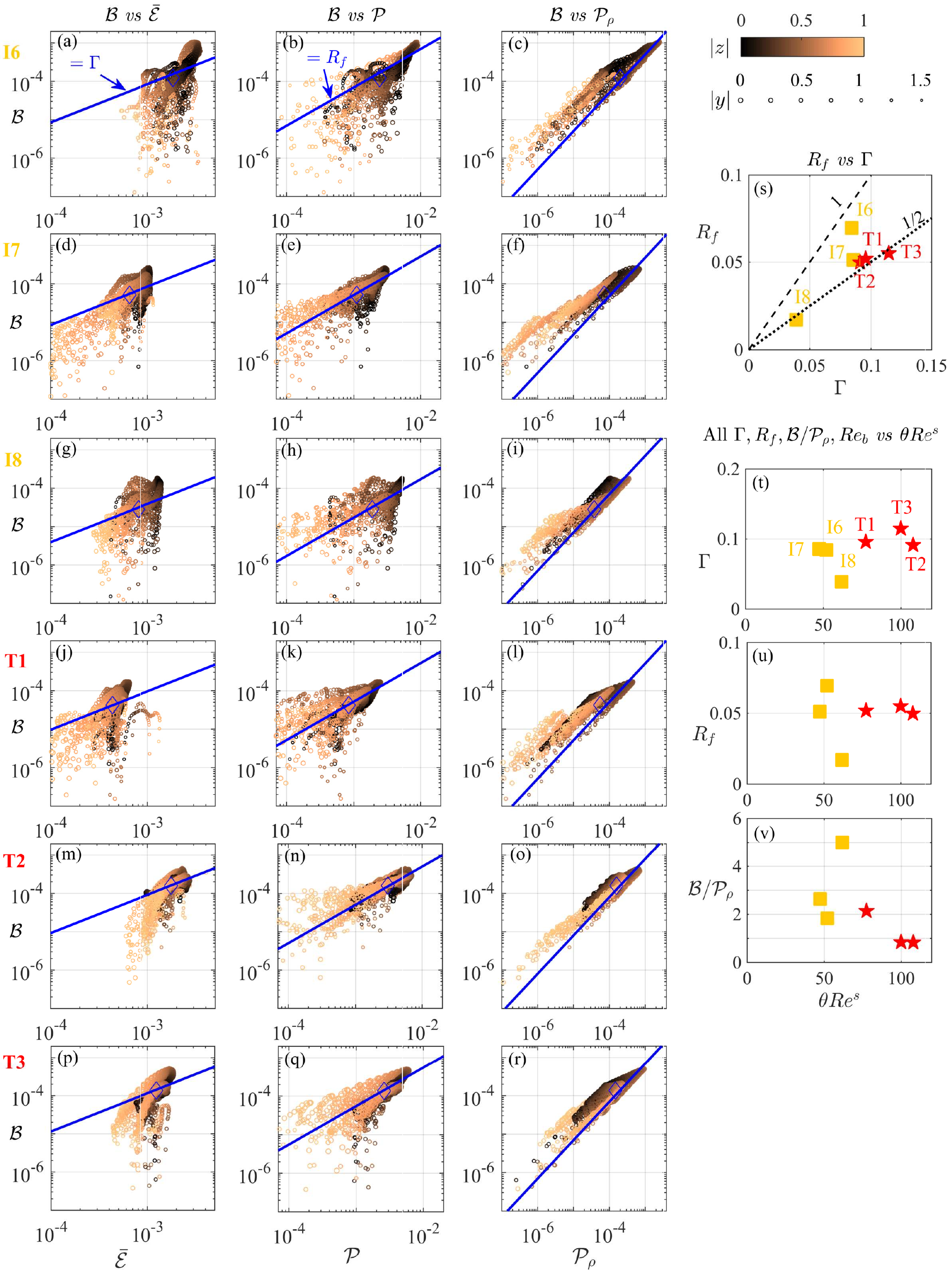}
    \caption{Flux coefficient, flux Richardson number, and $\mathcal{B}/\mathcal{P}_\rho$ ratio in data sets I6-T3 (top to bottom row). Clouds of $n_y n_z$ points of the  numerator $\mathcal{B}$  \emph{vs} the respective denominator:  $\bar{\mathcal{E}}$ (first column);  $\mathcal{P}$  (second column);   $\mathcal{P}_\rho$ (third column).  Log-log axes with identical vertical axis for all panels \emph{a-r}.  Symbol styles, and diamonds are as in figure~\ref{fig:parameterisation_1}. Right column:  $\Gamma, R_f,\mathcal{B}/\mathcal{P}_\rho$ \emph{vs} $\theta Re^s$ (values obtained from the linear fit or from the diamonds are indistinguishable).}
    \label{fig:parameterisation_2}
\end{figure}

In figure~\ref{fig:parameterisation_2}, we test the relations in \eqref{def-Gamma-Rf} with clouds of $n_y n_z$ data points (left three columns). The diamond coordinates are given by the numerator and denominator (volume averages) of \eqref{def-Gamma-Rf}; these were already plotted for all 16 data sets in
figure~\ref{fig:energetics_mean_fluxes}\emph{(j-l)}. Linear fits with enforced zero intercept are also shown in blue,  and provide the values for $\Gamma$, $R_f$, $\mathcal{B}/\mathcal{P}_\rho$  plotted in the rightmost column.

\begin{figure}
\centering
\includegraphics[width=0.95\textwidth]{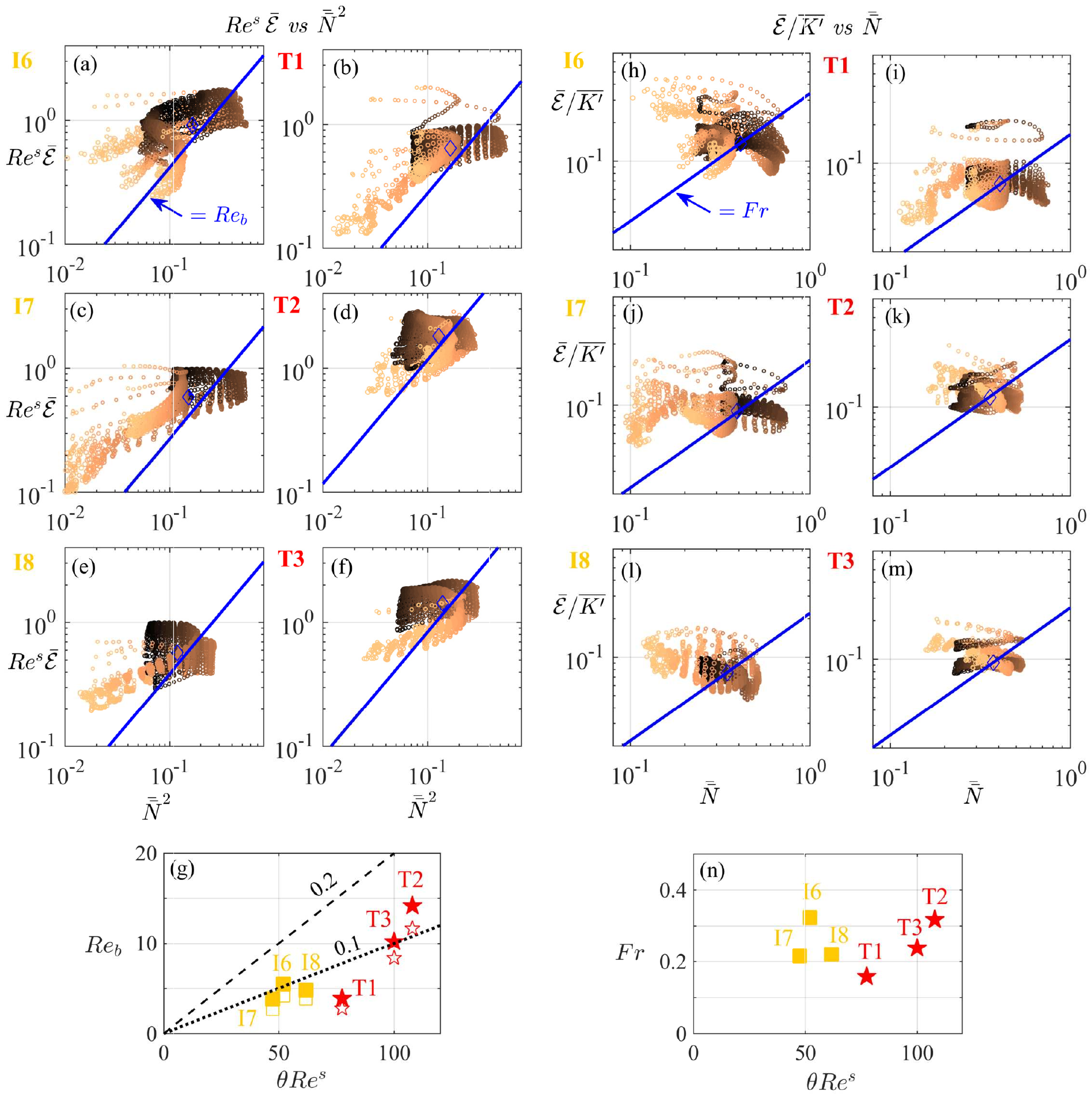}
    \caption{Buoyancy Reynolds number and turbulent Froude number in data sets I6-T3. Left two columns \emph{(a-f)}: numerator $Re^s\bar{\mathcal{E}}$ \emph{vs} denominator $\Nbb^2$. Right two columns  \emph{(h-m)}): numerator $\bar{\mathcal{E}}/\overline{K'}$ \emph{vs} denominator $\Nbb$. Log-log axes, symbol styles, and diamonds are as in figures~\ref{fig:parameterisation_1},\ref{fig:parameterisation_2}. Bottom row: \emph{(g)} $Re_b$  \emph{vs} $\theta Re^s$ to test \eqref{Reb_est1} (dashed line), showing with empty symbols the values obtained from the fit, and with full symbols the values obtained from the diamonds; \emph{(n)} $Fr$ \emph{vs} $\theta Re^s$ to test \eqref{def-Fr-2} (fit and diamonds values are indistinguishable). }
    \label{fig:parameterisation_3}
\end{figure}

First, focusing on the left three columns, we find that the clouds of the leftmost column have the largest spread, followed by those of the second column, and finally those of the third column, which are tighter around the fit. As a result, though the linear fits capture a clear trend, a constant $\Gamma$ is a relatively poor model (mean $r^2=0.30$), while  constant $R_f$ and $\mathcal{B}/\mathcal{P}_\rho$ are better models (mean $r^2=0.63$ and 0.66 respectively). Besides, the symbol colours or sizes do not reveal any clear pattern between this behaviour and the position $|z|,|y|$ within the shear layer. The diamonds generally lie very close to the fit, which means that our previous volume-averaged estimations of figure~\ref{fig:energetics_mean_fluxes}\emph{(j-l)} ($\Gamma\approx 0.1, R_f\approx 0.05, \langle \mathcal{B} \rangle /\langle \mathcal{P}_\rho\rangle \approx 1$) were good approximations.

Second, the values in the rightmost column  confirm indeed that $R_f \approx \Gamma/2$  (panel~\emph{s}, dotted line), which we recall is qualitatively sensible ($R_f<\Gamma$) but quantitatively inconsistent with  \eqref{eq-link-gamma-Rf}. Although this could be due to imperfections in the \cite{osborn_estimates_1980} balance \eqref{dKdt_0-1} shown in figure~\ref{fig:energetics_mean_fluxes}\emph{(e)} (our neglect of boundary fluxes), we believe it is more likely due to a systematic underestimation of $\langle \mathcal{E}\rangle$ (perhaps by a factor of 2), making  $\Gamma \approx R_f \rightarrow 0.05$  at high $\theta Re^s$ a perhaps more realistic asymptotic value than 0.1 (panels~\emph{t-u}). Moreover, we confirm that our data suggests $\mathcal{B}/\mathcal{P}_\rho \rightarrow 1$ at high $\theta Re^s$ (panel~\emph{v}), a necessary condition for the attractive `lock step' between $\mathcal{B}$ and $\chi$. 






\subsection{Buoyancy Reynolds number, turbulent Froude number}\label{sec:param-summary3}

In figure~\ref{fig:parameterisation_3}, we test the relations in \eqref{Reb_est0} (left two columns) and \eqref{def-Fr-1} (right two columns) with clouds of $n_y n_z$ data points, as in the previous two figures. Linear fits with enforced zero intercept are also shown in blue, and provide the values for $Re_b,Fr$ plotted in the bottom row against $\theta Re^s$ to test \eqref{Reb_est1} and \eqref{def-Fr-2} respectively.

First, we find that all clouds have a large spread around the fit. The linear fit in the left two columns (panels \emph{a-f}) captures a trend, especially the shape of the yellow cloud ($|z|\lesssim 1$ points at the edges of the shear layer), but a uniform $Re_b$ remains a poor model. The linear fit in the right two columns (panels \emph{h-m}) however even fails to capture the trend, arguing against a uniform $Fr$ model. These criticisms should be nuanced by the observation that the clouds in panels \emph{h-m} are very compact and span a very limited range (less than a decade in the horizontal and vertical axes). This limited range reveals an asymptotic tendency to uniform linear stratification (denominator), and to uniform dissipation frequency (numerator). The corresponding turbulent kinetic energy dissipation time scale  $(\bar{\mathcal{E}}/\overline{K'})^{-1} \rightarrow 10$ A.T.U.  (see the vertical co-ordinate of the diamonds), with some scatter in I6-T1 ($\approx 3-30$ A.T.U.), but much less scatter in T2-T3 ($\approx 5-12$ A.T.U.), suggesting some form of turbulent self-organisation.

Second, this observation allows us to deduce the asymptotic turbulent scaling $\langle K' \rangle \rightarrow 10 \langle \mathcal{E} \rangle \rightarrow 0.35\, \theta$ (using the scaling for $\langle \mathcal{E} \rangle$ in \eqref{estim-E-all}).  As a result, the parameterisation of eddy diffusivities $\nu_T,\kappa_T \propto Re^s \langle K' \rangle / \langle \Sbb \rangle$  proposed by \cite{reeuwijk_mixing_2019} in DNSs of inclined gravity currents yields $\nu_T,\kappa_T \propto \theta Re^s$  (using $\langle \Sbb \rangle \approx 1$, and our non-dimensionalisation of $\nu_T,\kappa_T$ by the molecular value $\nu$). This scaling appears compatible with our data in figure~\ref{fig:parameterisation_1}\emph{(u,v)}, although prefactors do not match.

Third, returning to figure~\ref{fig:parameterisation_3}, we find that, despite the spread of the clouds, the linear fits (blue lines) must (by construction) approximately go through the centre of mass of each cloud (blue diamonds), giving indistinguishable values of $Fr$ in panel \emph{n}. These data confirms our estimate \eqref{def-Fr-2} of an approximately constant $Fr\approx 0.3$. The values of $Re_b$ obtained from the  fit and from the diamonds are, however, slightly distinguishable, and shown using empty and full symbols respectively in panel \emph{g}. These data suggests an approximate scaling $Re_b \approx 0.1 \, \theta Re^s$ (dotted line), with volume-averaged values (full symbols) being consistently higher. This scaling is consistent with our estimate $Re_b \approx 0.2 \, \theta Re^s$ in \eqref{Reb_est1} (dashed line) if we again invoke the systematic underestimation of $\langle \mathcal{E}\rangle$ by a factor of 2.

\section{Conclusions}

In this Part 2 we presented some `advanced' properties of continuously-forced, shear-driven, stratified turbulence generated by exchange flow in a stratified inclined duct (SID) using the same 16 data sets and methodology as in Part 1. In \S~\ref{sec:energetics-equations} we introduced the evolution equations for the mean and turbulent kinetic energies and scalar variances which form the backbone of the remainder of the paper. We discussed approximate steady-state balances and compared them to the existing literature, and emphasised the SID-specific body forcing and boundary fluxes. Below we summarise the progress made on the three sets of questions raised in the end of \S~\ref{sec:intro}.

In \S~\ref{sec:energetics} we carried out the bulk of our turbulent energetics analysis. In \S~\ref{sec:energetics-time-vol-avgs} we first discussed the magnitude of all  time- and volume-averaged energy reservoirs, focusing on the variations in turbulent/mean and kinetic/scalar energy partitions in the Holmboe (H), intermittent (I) and turbulent (T) regimes. We then discussed the magnitude of all the key energy fluxes: the gravity forcing $\mathcal{F}$, the mean dissipation $\bar{\epsilon}$, the production of kinetic energy $\mathcal{P}$ and scalar variance $\mathcal{P}_\rho$, the buoyancy flux $\mathcal{B}$, the turbulent dissipation of kinetic energy $\mathcal{E}$ and scalar variance $\chi$, and the net advective flux of scalar variance $\Phi^{\bar{K}_\rho}$. We focused on critically assessing the validity of the simplified steady-state balances, carefully weighing our relative trust in theoretical expectations (based on conservation of energy) and in the accuracy of our measurements (limited at small scales, especially for $\mathcal{E}$ and $\chi$). We obtained empirical values for the flux ratios $\Gamma\equiv \mathcal{B}/\mathcal{E}$, $R_f\equiv \mathcal{B}/\mathcal{P}$ and $\mathcal{B}/\mathcal{P}_\rho$ and used these, together with higher-trust proxies (such as $\mathcal{F}$), with our physical understanding of hydraulic control ($\mathcal{E} \gg \,\bar{\epsilon}$ in $\TT$ flows), and with results from Part 1 ($Ri_b^s \approx 0.15$ in $\TT$ flows) to propose asymptotic (strongly turbulent) scaling laws for the rates and length scales of dissipation based on input parameters only (essentially $\mathcal{E},\chi \propto \theta$). We also highlighted the relevance of the product of parameters $\theta Re^s$ to measure the turbulence strength, measured by the square Frobenius norm of the turbulent strain rate tensor $||\mathsf{s}'||^2_F\equiv s_{ij}'s_{ij}' \approx 0.02 \theta Re^s$, (where $\theta\approx \tan \theta$ is the small tilt angle of the duct expressed in radians, and $Re^s$ is the shear layer, or `effective' Reynolds number). This importance of $\theta Re^s$ emerged in previous studies of the SID, and in the scaling of turbulent fractions in Part 1. It is consistent with the fact that an increasing large $\mathsf{s}'$ directly causes more extreme enstrophy through vortex stretching, noting that $||\mathsf{s}'||^2_F=  \sum_{i=1}^3\sigma_i^2 = \sum_{i=1}^3|\lambda_i|^2$ (where $\sigma_i,\lambda_i$ are respectively the three singular values and eigenvalues of $\mathsf{s}'$).

In \S~\ref{sec:energetics-spatiotemp} we investigated the spatio-temporal profiles of energy reservoirs and fluxes to articulate the specificity of SID turbulence. We discussed the characteristic vertical structure of the various turbulent sources and sinks across the shear layer, the spanwise effects, the temporal intermittency, and the potential importance of terms that we previously neglected for convenience (boundary fluxes) or by necessity (pressure terms) to accurately `close' the energy budgets.

In \S~\ref{sec:energetics-spectra} we examined the spectra of the turbulent kinetic and scalar energy,  first commenting on the decay exponent with the streamwise wavenumber, before breaking down all velocity components in all directions of space and time. We also compared two different methods to compute spectra from non-periodic, gridded experimental data (the direct Fourier transform and Welch's method).

In \S~\ref{sec:energetics-spectra-errors} we built on this spectral analysis to articulate six key limitations in the accuracy of our turbulent data in order to guide future technological developments (see Appendix~\ref{sec:appendix-limitations-spectra}). Some limitations are generic to experimental measurements (non-periodicity and finite-length, PIV/LIF filtering, resolution of turbulent length scales, finite-differentiation), while some are specific to our scanning system (volume reconstruction from successive planes and temporal aliasing). We also discussed the alternative computation of the (challenging) dissipation terms $\mathcal{E},\chi$ from model (ansatz) spectra and surrogate gradients, which raised the question of the anisotropy of our velocity data.

In \S~\ref{sec:anisotropy} we quantified this anisotropy. We first focused on the large-scale anisotropy of the Reynolds stress tensor  with a `Lumley triangle' mapping of all our data sets, explaining the generic tendency for strong prolate anisotropy (dominance of the streamwise velocity perturbation), pockets of oblate anisotropy in $\HH$ flows, followed by a more detailed analysis of the spatial structure of the individual tensor components underpinning $\mathcal{P}$. We then focused on the small-scale anisotropy of the 12 individual velocity gradients underpinning $\mathcal{E}$ (three longitudinal, six transverse, and three asymmetric terms). Assessing the relative accuracy of using each of them as a surrogate for $\mathcal{E}$ based on the assumption of isotropy (as is commonly done in field observations) suggested a tendency towards more isotropy with stronger turbulence,  quantified by the key product $\theta Re^s$.

In \S~\ref{sec:param} we tackled the parameterisation of turbulent energetics in our six most turbulent data sets. In \S~\ref{sec:param-background} we first sketched the hierarchy of simplified representations of the effects of mixing in terms of `direct' measures (eddy diffusivityes), `indirect' measures (flux coefficients $\Gamma,R_f$, mixing lengths), and key dynamical parameters (buoyancy Reynolds number $Re_b$, turbulent Froude number $Fr$). We then used our previous volume-averaged asymptotic (strongly turbulent) scaling laws to link these measures back to the only two `basic' flow parameters  $\theta$ and $Re^s$ that vary appreciably in this asymptotic regime, and we found that $Re_b \rightarrow 0.2\, \theta Re^s\approx 10-20$ and  $Fr \rightarrow 0.3$. This suggested that SID flows, as a result of hydraulic control, can be `vigorously' turbulent (predicting $Re_b \gg 30$, typically viewed as the threshold, for $\theta Re^s \gg 150$), while remaining strongly stratified ($Fr \ll 1$), at least provided $\theta$ remains small enough for the flow to remain largely horizontal (such that the mixing layer does not extend up to the vertical duct walls creating a mean streamwise stratification as in vertical exchange flows). These estimates allowed us to finally represent the expected relative order and separation of all the key length scales in SID turbulence (from the smallest to the largest: Batchelor, Kolmogorov, Corrsin, Ozmidov, Ellison/Thorpe, shear layer height, and duct size), highlighting in passing the current state and the desirable improvement of the PIV/LIF spatial resolution.

In \S\S~\ref{sec:param-summary1}-\ref{sec:param-summary3} we assessed \emph{a posteriori} the relevance of defining and using uniform values for these direct, indirect and parametric measures of mixing. Our data in $y-z$ revealed that most of these quantities were in fact non-uniform across the shear layer (to various degrees), which undermines the idealised models behind eddy diffusivities $\kappa_T,\nu_T$ (linear flux-gradient model), mixing lengths $L_m,L_\rho$ (quadratic flux-gradient model), flux parameters $\Gamma, R_f$ (linear relations between  $\mathcal{P},\mathcal{B},\mathcal{E}$), and dynamic parameters $Re_b,Fr$ (uniform ratios of length scale and time scales). These significant reservations aside, we found that the earlier volume-averaged estimates $\Gamma \approx 0.1$ and $R_f \approx 0.05$ were  representative fits of the underlying clouds of data, although we argued that our current underestimation of $\mathcal{E}$ makes $\Gamma \approx 0.05$ a more plausible value, i.e. a 5\,\% `tax'  confidently below the 20\,\% `tax'  found in most of the  literature. We confirmed that $Re_b \rightarrow 0.2 \theta Re^s$ (invoking the same underestimation of $\mathcal{E}$), that $Fr \rightarrow 0.3$, and that the turbulent Prandtl number (ratio of eddy diffusivities) $Pr_T \approx 0.15/R_f\rightarrow 3$ (where $0.15$ is the `equilibrium Richardson number' found in Part 1), which is confidently above 1 and representative of strongly-stratified turbulence.
We also confirmed that asymptotically $\mathcal{B}/\mathcal{P}_\rho \rightarrow 1$ as expected under approximately linear stratification, i.e. that $\mathcal{B},\mathcal{P}_\rho,\chi,\Phi^{\bar{K}_\rho}$ tend to a balance (or `lock step'). Under such a conceptually attractive lock step, $\chi$ becomes approximately equivalent to the rate of irreversible mixing (destruction of available potential energy), and $\Gamma$ becomes equivalent to $\chi/\mathcal{E}$, the `real taxation rate' of stratification.

\vspace{0.5cm}

\noindent \textbf{Acknowledgements}

\vspace{0.2cm}
AL is supported by an Early Career Fellowship funded by the Leverhulme Trust and the Isaac Newton Trust. We also acknowledge past funding from EPSRC under the Programme Grant EP/K034529/1 `Mathematical Underpinnings of Stratified Turbulence' (MUST) and current funding from the European Research Council (ERC) under the European Union's Horizon 2020 research and innovation Grant No 742480  `Stratified Turbulence And Mixing Processes' (STAMP). Finally, we are grateful for the invaluable experimental support and expertise of Stuart Dalziel, Jamie Partridge and the technicians of the G. K. Batchelor Laboratory.  

\vspace{0.5cm}

\noindent \textbf{Declaration of Interests}

\vspace{0.2cm}

 The authors report no conflict of interest.

\appendix{

\section{Computation of energy spectra} \label{sec:Appendix-spectra}

In this appendix we define and explain how we computed the energy spectral densities introduced in \eqref{def-spectral-densities-in-x} and plotted in figures~\ref{fig:spectra_x_all}-\ref{fig:spectra_xyzt} (which are not standard in the literature).

\subsection{Continuous definitions}

We define the spectral density $E^x_{\psi'}$ along $x$ of any perturbation variable $\psi'$ in a continuous sense (using integrals) as follows:
\begin{subeqnarray} \label{continuous-spectrum-1}
\langle \psi'^2 \rangle  &\equiv&  \frac{1}{8 L_x L_y L_z L_t} \int_{0}^{L_t}  \int_{-1}^{1}  \int_{-L_y}^{L_y}  \Big( \int_{0}^{2L_x} \psi'^2 \d x \Big) \d y \d z \d t \quad \text{(definition)}  \quad  \slabel{continuous-spectrum-2} \\
    &=& \frac{1}{8 L_x L_y L_z L_t} \int_{0}^{L_t}  \int_{-1}^{1}  \int_{-L_y}^{L_y} \Big( \frac{1}{2\pi} \int_0^{k_x,\, max}|\widehat{\psi'}|^2 \d k_x \Big)  \d y \d z \d t \quad \text{(Parseval)}  \qquad  \quad \slabel{continuous-spectrum-3}\\
&=&  \int_{0}^{k_x,\, max} \frac{1}{4 \pi  L_x}  \Big( \frac{1}{4 L_y L_z L_t} \int_{0}^{L_t} \int_{-1}^{1}  \int_{-L_y}^{L_y}  |\widehat{\psi'}|^2 \d y \d z \d t \Big) \d k_x \  \text{(re-arranging)}  \qquad \quad \slabel{continuous-spectrum-4}\\
&\equiv& \int_{0}^{k_x,\, max} E^x_{\psi'} \d k_x \quad \text{(definition \eqref{def-spectral-densities-in-x})} \  \implies \ E^x_{\psi'}(k_x) \equiv \frac{1}{4\pi L_x} \langle |\widehat{\psi'}|^2 \rangle_{y,z,t}.\slabel{def-Expsi}
\end{subeqnarray}
We used the definitions for averages and fluctuations in  Part 1, \eqref{def-spectral-densities-in-x}, and Parseval's theorem stating that the total energy of $\psi'$ along $x$ is conserved in its Fourier transform $\widehat{\psi'}(k_x,y,z,t)$.

Applying the above definition \eqref{def-Expsi} to $\psi'=u',v',w',\rho'$, we find (note the $1/2$ factor for energies $(u'^2)/2$, etc):
\begin{equation}
\begin{gathered}
E^x_{u'} \equiv \frac{1}{8\pi L_x} \langle |\widehat{u'}|^2 \rangle_{y,z,t}, \quad  E^x_{v'} \equiv \frac{1}{8\pi L_x} \langle |\widehat{v'}|^2 \rangle_{y,z,t} ,\quad E^x_{w'} \equiv \frac{1}{8\pi L_x} \langle |\widehat{w'}|^2 \rangle_{y,z,t}, \\ 
E^x_{K'} \equiv E^x_{u'}+E^x_{v'}+E^x_{w'}, \qquad E^x_{K'_\rho} \equiv   \frac{Ri_b^s }{8\pi L_x} \langle |\widehat{\rho'}|^2 \rangle_{y,z,t}
\end{gathered}
\end{equation}
All of the above can be extended naturally from $x$ to $y,z,t$.

\subsection{Discrete definitions}\label{sec:Appendix-spectra-discrete-def}

Here we provide the exact expressions for the discrete analogue of the above definitions used in numerical computations on our gridded data.

Consider a perturbation signal $\psi'_{qrst}$ with discrete grid values are indexed  by  $q=1,2,\ldots,n_x$, (and similarly with $n_y$, $n_z$, $n_t$ grid points in $r,s,t$ respectively). The discrete analogue of of \eqref{continuous-spectrum-2} for the time- and volume-averaged energy is
\begin{subeqnarray} \label{discrete-avg}
\langle \psi'^2 \rangle  =  \frac{1}{n_x n_y n_z n_t} \sum_{t=1}^{n_t} \sum_{s=1}^{n_z}  \sum_{r=1}^{n_y}  \sum_{q=1}^{n_x} (\psi'_{qrst})^2 .
\end{subeqnarray}
Note that $\Delta x/(2L_x)=1/n_x$, $\Delta y/(2L_y)=1/n_y$, etc. We used simple sums (rectangular integration) here and in all computations of energy spectra involving discrete Fourier transforms in order to satisfy Parseval's conservation of energy. However, in the remainder of the paper, we used trapezoidal integration to compute averages for better accuracy.

The discrete energy spectral densities of $\psi'$ along $x,y,z,t$ are defined respectively as $E^x_m,E^y_n,E^z_o,E^t_p$ (the subscript $\psi'$ is implicit and omitted for clarity), with discrete grid values indexed by $m,n,o,p$ in the wavenumber/frequency space $k_x,k_y,k_z,\omega$, where:
\begin{subeqnarray} \label{discrete-spectrum-1}
\langle \psi'^2 \rangle  &=& \sum_{m=1}^{n_m} E^x_{m} \Delta k_x  = \sum_{m=1}^{n_m} \underbrace{\Big( \frac{1}{n_y n_z n_t} \sum_{t=1}^{n_t} \sum_{s=1}^{n_z}  \sum_{r=1}^{n_y}  E^x_{mrst} \Big) }_{\equiv E^x_m} \Delta k_x ,\\
&=& \sum_{n=1}^{n_n} E^y_{n} \Delta k_y  = \sum_{n=1}^{n_n} \underbrace{\Big( \frac{1}{n_x n_z n_t} \sum_{t=1}^{n_t} \sum_{s=1}^{n_z}  \sum_{q=1}^{n_x}  E^y_{qnst} \Big)}_{\equiv E^y_n} \Delta k_y ,\\
&=& \sum_{o=1}^{n_o} E^z_{o} \Delta k_z  = \sum_{o=1}^{n_o} \underbrace{\Big( \frac{1}{n_x n_y n_t} \sum_{t=1}^{n_t} \sum_{r=1}^{n_y}  \sum_{q=1}^{n_x}  E^z_{qrot} \Big)}_{\equiv E^z_o} \Delta k_z, \\
&=& \sum_{p=1}^{n_p} E^t_{p} \Delta \omega  = \sum_{p=1}^{n_p} \underbrace{\Big( \frac{1}{n_x n_y n_z} \sum_{s=1}^{n_z} \sum_{r=1}^{n_y}  \sum_{q=1}^{n_x}  E^t_{qrsp} \Big)}_{\equiv E^t_p} \Delta \omega.
\end{subeqnarray}
In each line, the first equality is our definition of spectral energy density (as in the first equality of \eqref{def-Expsi}),  the second equality comes from Parseval's theorem (as in  \eqref{continuous-spectrum-3}-\eqref{continuous-spectrum-4}), and 
\begin{subeqnarray}\label{discrete-spectrum-2}
E^x_{mrst} &\equiv&  \frac{\Delta x}{\pi n_x(1+\delta_{m,1}+\delta_{m,n_m})} |\widehat{\psi'}_{mrst}|^2, \qquad m=1,\ldots,n_m\equiv \frac{n_x}{2}+1, \slabel{discrete-spectrum-2-a} \\
E^y_{qnst} &\equiv&  \frac{\Delta y}{\pi n_y(1\,+ \,\delta_{n,1}\, + \, \delta_{n,n_n})} |\widehat{\psi'}_{qnst}|^2,\qquad n=1,\ldots,n_n\equiv \frac{n_y}{2}+1, \\
E^z_{qrot} &\equiv&  \frac{\Delta z}{\pi n_z(1\,+\,\delta_{o,1}\,+\,\delta_{o,n_o})} \, |\widehat{\psi'}_{qrot}|^2 ,\, \qquad o=1,\ldots,n_o\equiv \frac{n_z}{2}+1,\\
E^t_{qrsp} &\equiv&  \frac{\Delta t}{\pi n_t(1\,+\,\delta_{t,1}\,+\,\delta_{t,n_t})} \,\, \, |\widehat{\psi'}_{qrsp}|^2 , \, \qquad p=1,\ldots,n_p\equiv \frac{n_t}{2}+1,
\end{subeqnarray}
where the $|\widehat{\psi'}_{mrst}|^2$ are the square moduli of the one-dimensional discrete Fourier transforms (DFTs) of $\psi'_{qrst}$ along $x$:
\begin{equation}\label{discrete-FT}
    \widehat{\psi'}_{mrst} \equiv \sum_{q=1}^{n_x} \psi'_{qrst} \, e^{-\frac{2i\pi}{n_x} (q-1)(m-1)}, \qquad k_{x,\,m} \equiv (m-1) \Delta k_x \equiv (m-1) \frac{\pi}{L_x},
\end{equation}
and similarly along $y,z,t$. In \eqref{discrete-spectrum-2}, the Kronecker $\delta$ (e.g. $\delta_{m,1}=1$ if $m=1$, and 0 otherwise) is used because we consider the positive (one-sided) spectrum of a real signal, resulting in energy being counted twice at the $0$ and maximum (Nyquist) frequencies. The normalisation constant in $\Delta x/(\pi n_x) = 2L_x/(\pi n_x^2)$ is consistent with Matlab's `fft' function convention to attach the $1/n_x$ normalisation factor to the inverse transform (rather than to the forward transform). The density of $u'^2/2$ is given by replacing $|\widehat{\psi'}|^2$ by $|\widehat{u'}|^2/2$ in  \eqref{discrete-spectrum-2}, etc. 

For more details about computing energy spectra from gridded data with correct normalisation, see \cite{durran_practical_2017} (\S~2).

\subsection{Energy at zero wavenumber/frequency} \label{sec:Appendix-spectra-energy-k0}

Reverting back to continuous variables for simplicity, the energy spectral density of $\psi'$ at zero wavenumber/frequency $(k_x,k_y,k_z,\omega)=(0,0,0,0)$ is:
\begin{subeqnarray}\label{non-zero-spectrum-at-0-freq}
E^x_{\psi'}(k_x=0) &=& \frac{2L_x}{\pi} \langle \langle \psi' \rangle_x^2 \rangle_{y,z,t} \neq 0 , \\
E^y_{\psi'}(k_y=0) &=&  \frac{2L_y}{\pi} \langle \langle \psi' \rangle_y^2 \rangle_{x,z,t} \neq 0, \\
E^z_{\psi'}(k_z=0) &=&  \frac{2L_z}{\pi} \langle \langle \psi' \rangle_z^2 \rangle_{x,y,t} \neq 0,\\
E^t_{\psi'}(\omega=0) &=&   \frac{L_t}{\pi} \langle \langle \psi' \rangle_t^2 \rangle_{x,y,z} \neq 0.
\end{subeqnarray}
We used \eqref{def-Expsi} and the fact that by definition of the Fourier transform  along $x$
\begin{equation}
    |\widehat{\psi'}(k_x=0,y,z,t)|^2 \equiv \Big(\int_0^{2L_x} \psi'(x,y,z,t) \d x \Big)^2 = (2L_x\langle \psi' \rangle_x)^2,
\end{equation}
and similarly along $y,z,t$.

The values in \eqref{non-zero-spectrum-at-0-freq} are essentially mean variances along $x,y,z,t$, respectively, that are generally non-zero because our data is four-dimensional, and our definition of $\psi'\equiv \psi-\langle \psi \rangle_{x,t}$ does not guarantee that $\langle \psi'\rangle_{\xi}^2$ averages to zero for any single coordinate $\xi$. This is in contrast with typical practice with one-dimensional data, where perturbations are defined as $\psi'(\xi) \equiv \psi - \langle \psi \rangle_{\xi}$ (such that $\langle \psi' \rangle_{\xi}=0$), resulting in $E^{\xi}_{\psi'}(k_{\xi}=0)=0$.

\subsection{Welch's method} \label{sec:Appendix-spectra-welch}
 
Welch's method \citep{welch_use_1967} is a non-parametric estimator of the energy spectral density of a signal that  minimises both spectral leakage (Gibbs phenomenon) caused by non-periodicity of the data (edge discontinuities) and measurement noise. 

To render the data periodic, a `Hamming' window function is applied (tapering to zero at the edges). Windowing reduces spectral leakage at the expense of resolution in frequency space, because it effectively shortens the usable length of the original signal. Windowing alone results in a loss of information by giving more importance to the central portion of the signal.

To mitigate this loss and give more equal importance to the whole signal, the signal is instead divided into a series of overlapping segments of equal length, windowing is applied to each individual segments, and Welch's spectral density is computed by averaging the square modulus of each individual DFTs (we used Matlab `pwelch' function with eight segments and 50~\% overlap between segments). As segmentation reduces resolution in frequency space, it remains attractive only if the signal is long enough for frequency resolution to be a lesser concern (this is the case for us in $x,t$ because typically $n_x,n_t\gg 100$, but not in $y,z$, explaining why we do not plot Welch's method in figure~\ref{fig:spectra_xyzt}\emph{d-i}). 

Welch's segmentation and averaging also have the key benefit of reducing experimental measurement noise (the variance of the noise in Welch's estimated spectrum reduces in proportion to the number of segments). For more details, see \cite{smith_digital_2003} (Chap.~9).

\subsection{Note on three-dimensional Fourier transforms}\label{sec:Appendix-spectra-3DFT}

Here we explain why we defined energy spectral densities using one-dimensional rather than three-dimensional Fourier transforms in $x,y,z$.

Theoretical and numerical studies on homogeneous isotropic turbulence usually consider the one-dimensional energy spectrum $\langle K' \rangle \equiv \int_0^\infty E(k) \d k$, where $k\equiv |\mathbf{k}| = (k_x^2+k_y^2+k_z^2)^{1/2}$ (e.g. \citealp{batchelor_theory_1953} eq. (3.1.6)), obtained by averaging the three-dimensional Fourier transforms $|\widehat{u'}(\mathbf{k})|^2,|\widehat{v'}(\mathbf{k})|^2,|\widehat{w'}(\mathbf{k})|^2$ on spherical shells of equal $k$. Although formally attractive (e.g. dissipation is obtained simply as $\langle \mathcal{E} \rangle= \int_0^\infty  k^2 \, E(k) \d k$), this formulation is of limited use and impractical for our data.

First, our flows are inhomogeneous and anisotropic, at least at the scales that can be resolved (see \S~\ref{sec:anisotropy}). We can neither treat all directions equally nor use the attractive formula for the dissipation  $\int k^2 \, E(k)$.

Second, our data is far from being triply-periodic, and is given on a discrete grid with different spacings $\Delta x, \Delta y, \Delta z$ and domain lengths $L_x,L_y,L_z$. To our knowledge, it is impossible to compute a sensible and energy-preserving one-dimensional shell-average of a three-dimensional Fourier transform performed on a wavenumber grid having vastly different $\Delta k_x,\Delta k_y,\Delta k_z$ and $k_{x, \, max},k_{y, \, max},k_{z, \, max}$. Even with a more ideal domain and grid, the shell-averaging of gridded data creates inherent noise. This noise can be reduced by some \emph{ad hoc} techniques, but these techniques  do not conserve energy (\citealp{durran_practical_2017}, \S~3).

\section{Limitations of our energetics data}\label{sec:appendix-limitations-spectra}

In this appendix we complement the discussion in \S~\ref{sec:energetics-spectra-errors} by providing further information regarding the five key current limitations in our computation of energetics, based on insights derived from spectral data in \S~\ref{sec:energetics-spectra}. 

\subsection{Non-periodic and finite-length data} 

As mentioned in \S~\ref{sec:energetics-spectra} and in Appendix~\ref{sec:Appendix-spectra-welch} non-periodic and finite-length data causes the high-wavenumber content of our spectra to be polluted by spectral leakage. This is particularly true in $y$ and $z$ due to the limited number of data points (domain length and resolution), where Welch's method is inapplicable.

\subsection{PIV and LIF filtering}

First, the cross-correlation of PIV across interrogation windows (IWs)  effectively convolves the underlying `real' velocity field with a square filtering kernel of size $\ell_{IW}$ in $x$ and $z$. This filtering can -- in principle -- be corrected for, by multiplying the energy densities $E^x_{K'}$, $E^z_{K'}$ by the inverse energy density of the filtering kernel $\propto (\ell_{IW} k_x/2)^2/\sin^2(\ell_{IW} k_x/2)$ (and similarly in $z$), as proposed in \cite{xu_accurate_2013} in their \S~4.2. However this rescaling function is singular at the IW wavenumber $2\pi/\ell_{IW}$, and thus requires a Nyquist wavenumber $k_{x,\,max} \equiv \pi/dx \le 2\pi/\ell_{IW}$, i.e. a grid spacing $dx\ge \ell_{IW}/2$, corresponding to a requirement of $\le 50$\,\% overlap between IWs. However, $>50$\,\% overlap (oversampling) is unfortunately common and practical in PIV (our data uses $62$\,\% overlap). 

Second, LIF also effectively averages the density field to pixel resolution, and we further low-pass filtered these data to remove various sources of noise (e.g. due to spurious rays caused by dust in the optical path of the laser sheet), before sub-sampling them to the lower-resolution PIV grid $dx,dz$ for convenience. Such steps could be avoided or improved (and our spectra of $E_{K'_\rho}$ could have indeed been given up to higher Nyquist wavenumbers $k_{x,\,max}, k_{z, \max}$ in figures~\ref{fig:spectra_x_all}-\ref{fig:spectra_xyzt}). However, we verified that this would yield very limited practical benefits given the daunting separation between the Batchelor and Kolmogorov  scales ($\ell_B \approx  \ell_K/25$).

Third, both our PIV and our LIF data are inherently averaged in $y$ across the thickness of the laser sheet (the filterning kernel depends on the poorly-known laser sheet intensity $y$ profile). We performed the experiments with a spacing $dy$ approximately equal to the mean laser sheet thickness to avoid $>50$\,\% overlap in $y$ (oversampling), but uncertainties remain. 

Fourth, we recall that the spanwise component $v'$ seems partially contaminated with medium- to small-scale noise along $x$, presumably as a result of slight and poorly-understood errors in the delicate stereo-PIV computation of this out-of-plane velocity component.

\subsection{Resolution of the Kolmogorov and Batchelor length scales} 

The rescaling mentioned above to correct for PIV and LIF filtering is only expected to significantly improve measures of energy and dissipation on properly-sampled data if the Nyquist wavenumbers $k_{x,\,max},k_{y,\,max},k_{z,\,max}$ are comparable with $k_K = 2\pi/\ell_K$ (for PIV) and $k_B = 2\pi/\ell_B$ (for LIF), where $\ell_K, \ell_B$ are defined and estimated in \eqref{def-lK}-\eqref{def-lB}. Although the Kolmogorov wavenumber is within reach in $x,z$ (see \S~\ref{sec:energetics-Kolmogorov}), and potentially in $y$ with improvements in the apparatus, the Batchelor wavenumber will likely always remain out of reach at $Pr=700$. Note that measurements in temperature-stratified flows at $Pr=7$ are unfortunately impractical, because of the inability to have a uniform refractive index.

\subsection{Volume reconstruction and spanwise distortion}

Our three-dimensional volumetric data are reconstructed in $y$ by aggregating successive $x-z$ planes obtained at slightly different times (it takes a time $\Delta t$ to scan from one duct wall to another $-1 \le y^h \le 1$).
The resulting spanwise distortion of turbulent structures could (and probably does) affect energy estimates. It appears tempting to correct for this distortion using G. I. Taylor's hypothesis that turbulent fluctuations $\uu',\rho'$ are `frozen' and advected by the mean flow $\bar{u}(y,z)$. This would require a non-trivial $x$-coordinate map $X(x,y,z,t) \equiv x -  \bar{u}(y_i,z)(t_i-t)$, where $t_i-t$ is the time difference between  the exact time at which plane $y_i$ was captured and the mean time at which each reconstructed volume is given. However, this does not appear viable since it would cause further spurious distortions (because Taylor's hypothesis is questionable with inhomogenous flows $\bar{u}(y,z)$), and it would further reduce the spanwise resolution of our data (because of the lack of $x$ periodicity, data within a distance $\max_{y,z}|\bar{u}|\Delta t L_y/2 \approx \Delta t$ of each end would be lost, which can be considerable).

\subsection{Temporal resolution and aliasing} 
Our scanning time step $\Delta t$ between volumes is decades higher than the smallest dynamically-relevant turbulent timescale, i.e. turbulent energy is contained well above our Nyquist frequency $\omega_{max}$. This causes aliasing of temporal spectra, whereby unresolved high-frequency energy is incorrectly mirrored into resolved low-frequency energy (\citealp{smith_digital_2003}, pp.~39-45; \citealp{springer_handbook_2017}, \S~22.1). Note that this effect is only expected in temporal spectra (which may or may not be of interest) due to sampling in $t$ being achieved by very short laser pulse duration (for LIF) and laser pulse separation (for PIV), whereas in $x,y,z$ the filtering/averaging effects of PIV/LIF dominate. 

\subsection{Finite differentiation}  Direct estimations of $\mathcal{E},\chi$ by finite differentiation in physical space are prone to further errors, because standard finite-difference operators effectively convolve the data by a set of offset rectangular window functions whose spectra have high-amplitude side-lobes. Although more advanced finite-difference schemes with improved (smoother)  properties exist, they nevertheless inevitably amplify the high-wavenumber inaccuracies of the original signal.

}

\bibliographystyle{jfm}

\bibliography{AL_references_2020_12.bib}

\clearpage

\tableofcontents

\end{document}